# Extended Short-Wave Infrared Absorption in Group IV Nanowire Arrays

A. Attiaoui,[1,*] É. Bouthillier,[1] G. Daligou,[1] A. Kumar,[1] S. Assali,[1] and O. Moutanabbir[1,*]

[1] Department of Engineering Physics, École Polytechnique de Montréal, C. P. 6079, Succ. Centre-Ville, Montréal, Québec H3C 3A7, Canada

**ABSTRACT:**

Engineering light absorption in the extended short-wave infrared (e-SWIR) range using scalable materials is a long-sought-after capability that is crucial to implement cost-effective and high-performance sensing and imaging technologies. Herein, we demonstrate enhanced, tunable e-SWIR absorption using silicon-integrated platforms consisting of ordered arrays of metastable GeSn nanowires with Sn content reaching 9 at.% and variable diameters. Detailed simulations were combined with experimental analyses to systematically investigate light-GeSn nanowire interactions to tailor and optimize the nanowire array geometrical parameters and the corresponding optical response. The diameter-dependent leaky mode resonance peaks are theoretically predicted and experimentally confirmed with a tunable wavelength from 1.5 to 2.2 µm. A three-fold enhancement in the absorption with respect to GeSn layers at 2.1 µm was achieved using nanowires with a diameter of 325 nm. Finite difference time domain simulations unraveled the underlying mechanisms of the e-SWIR enhanced absorption. Coupling of the $HE_{11}$ and $HE_{12}$ resonant modes to nanowires is observed at diameters above 325 nm, while at smaller diameters and longer wavelengths the $HE_{11}$ mode is guided into the underlying Ge layer. The presence of tapering in NWs further extends the absorption range while minimizing reflection. This ability to engineer and enhance e-SWIR absorption lays the groundwork to implement novel photonic devices exploiting all-group IV platforms.

**KEYWORDS: extended short-wave infrared, absorption, nanowire array, germanium-tin, silicon photonics**

E-SWIR (~1.4–2.5 μm) responds primarily to reflected light from objects and can penetrates fog and smog, thus enabling imaging and sensing through scattering media.[1–3] In principle, e-SWIR devices can passively operate even in the dark as the nightglow from the upper atmosphere provides a natural illumination in this wavelength range. Additionally, e-SWIR sensors can also be employed for thermal imaging of objects at temperatures above 150 °C,[4] as well as for the spectral identification of substances sharing O-H, C-H and N-H bonds[5,6] exploiting the strong absorbance related to molecular vibrations. Deciphering these molecular fingerprints is highly relevant for chemical and biomedical technologies. However, the high production cost and the limited spectral tunability have been a major obstacle facing the widespread adoption of e-SWIR imaging and sensing capabilities.[7] Indeed, current e-SWIR detectors are based predominantly on quantum well IR photodetectors, InGaAs, InSb, or HgCdTe.[8,9] As a matter of fact, these materials are prohibitively expensive which translates into a limited imaging and sensing array size and an overall cost of a megapixel e-SWIR sensor that can exceed tens of thousands dollars.[7]

An attractive alternative consists of exploiting the emerging group IV germanium-tin (GeSn) semiconductors, which exhibit a bandgap energy that can be tuned to cover the e-SWIR range and beyond. Indeed, the bandgap energy in these semiconductors shrinks from, for instance, 1.72 μm (0.72 eV) to 8.0 μm (0.15 eV) as Sn content from 5 at.% to 30 at.% in absence of any lattice strain.[10–15] However, these compositions are substantially higher than the ~1 at.% solubility of Sn in Ge. To reach this compositional range, nonequilibrium growth protocols were developed to prevent phase separation and avoid Sn segregation and material degradation.[12,16–21] Interestingly, GeSn layers can be grown epitaxially on silicon wafers thus allowing a full compatibility with complementary metal-oxide-semiconductor (CMOS) processing needed for a full exploitation of the current microelectronic and photonic technologies. This would yield production in a high-volume with repeatability, uniformity, and cost-effectiveness using standard design flows.

Recognizing the potential of GeSn as an effective building block for CMOS-compatible sensing and imaging devices, tremendous efforts have been recently expended to integrate this material system in design and fabrication of a variety of optoelectronic and photonic devices.[22–26] This demonstrated capability of GeSn as a versatile, silicon-compatible e-SWIR material can be further enhanced by exploiting the nanowire (NW) geometry as an additional degree of freedom to engineer and control the light-matter interaction. Indeed, NWs exhibit excellent anti-reflection properties over an extended wavelength range [27] and are a medium with a proven enhanced absorption and efficiency in solar cells and photodetectors.[27–29] Furthermore, NWs also provide the control over the directionality and polarization of the emitted light by simply tuning the diameter and tapering without any external optics.[30–33] The enhanced strain relaxation along the NW radial direction enabled the development of both axial and radial bottom-up GeSn NW heterostructures with tunable room-temperature direct band gap emission.[34–38] Herein, we demonstrate the use of GeSn NW arrays on a Si wafer to achieve a tunable and enhanced e-SWIR absorption. Leaky mode resonance peaks with a variable wavelength from 1.5 to 2.2 µm are theoretically predicted and experimentally achieved in GeSn NW arrays at a Sn content reaching ~9 at.%. The enhanced e-SWIR absorption is driven by the coupling of $HE_{11}$ mode to NWs at diameters above 325 nm, while at smaller diameters the $HE_{11}$ mode is guided into the underlying Ge layer.

# RESULTS AND DISCUSSION

**GeSn epitaxy and NW array fabrication.** The NW arrays were dry-etched from epitaxially grown 1.1 μm-thick GeSn layer on a 1.0 μm-thick Ge-on-Si virtual substrate (Ge-VS) (see Methods for details). The cross-sectional transmission electron micrograph (TEM) and the energy dispersive X-ray spectroscopy (EDS) compositional profile are shown in Figure 1a. Misfit dislocations are mainly visible at GeSn/Ge interface as a result of the lattice-mismatched growth. The EDS profile shows a graded composition towards the surface of GeSn resulting from the strain relaxation during growth.[20] To evaluate both Sn content and lattice strain, X-ray diffraction (XRD) reciprocal space mapping (RSM) measurements around the asymmetrical (224) reflection were carried out (Figure 1b). The small tensile strain of $\varepsilon_{||} \sim 0.16$ % in Ge layer is thermal mismatch-induced during cyclic annealing prior to GeSn growth. The incorporation of Sn across the graded GeSn layer increases from 6.7 to 9.2 at.% toward the surface, while the compressive strain increases from $\varepsilon_{||} \sim -0.2$ % to $\sim -0.4$ %. Next, GeSn NW arrays with variable diameters were fabricated using a $Cl_2$-based reactive-ion etching (RIE) process (see Methods for details). This process yields slightly tapered NWs. Tapering is expected to improve light collection and absorption, as discussed below.

Based on the systematic simulations discussed below, four different tapered NW arrays were fabricated with diameters (top/bottom) of 175/300 nm (S1), 200/325 nm (S2), 325/450 nm (S3), and 375/550 nm (S4), as shown in the Scanning Electron Microscopy (SEM) images in Figure 1c. A pitch of 1.1 μm was used for S1, S2, and S3, while a pitch of 1.6 μm was used for S4 to account for the larger NW diameter. Few (thinner) parasitic NWs are visible in S1 and S2, while at the largest diameter S4 the NW top-surface seems to be not perfectly flat. This subtle morphological feature has no effect of on the optical response of these NWs. NW tapering, as exemplified in Figure 1d, originates most likely from parasitic etching of the resist on top and sidewall diffusion of ionic species during the RIE process. The height of NWs, ~1.3 µm, is slightly larger than the as-grown GeSn layer thickness due to the partial etching of Ge layer, as discussed in Supporting Information S1. In addition, the compressive strain $\varepsilon_{||} \sim -0.4$ % in the as-grown GeSn layer has been fully relaxed in NWs due to the formation of free surfaces (Supporting Information S1). The geometrical parameters defining the four arrays are summarized in Table 1. Before discussing the experimental measurements, the optical properties of these arrays are first evaluated based on simulations.

**Numerical simulations of the absorption in GeSn NW arrays.** Three-dimensional finite-difference time-domain (FDTD) simulations were performed using the Lumerical© software package to assess the effects of the array parameters on light absorption. For precise simulations, we first developed the optical model for as-grown GeSn layers. In this regard, the complex dielectric function $\tilde{\varepsilon} = \varepsilon_1 + i\varepsilon_2$ in the 0.9-2.5 µm spectral range was estimated from Spectroscopic Ellipsometry (SE) measurements (Supporting Information S2). The obtained dielectric function was used as input in FDTD simulations. The optical response of triangular arrays was investigated at a near-normal incidence (5°). Figure 2a depicts the relevant geometrical parameters used in these simulations, namely the diameter *d*, the pitch *u* and the height *H*. For simplicity, cylindrical NWs with a uniform composition of 9 at.% are considered in the simulations. NW tapering will also be addressed as discussed below. Figure 2b exhibits the simulated absorptance (*A*) map for a triangular array of 1.1 µm-long GeSn NWs at a wavelength of 2 µm as a function of *d* and *u* up to 1.5 µm and 2.2 µm, respectively. Multiple leaky mode resonances (LMRs) are visible at different diameters, while the pitch has a subtle effect on the absorption. Two sharp resonances are observed at diameters of 350 nm and 850 nm, while the weaker dependence on the pitch is highlighted in the inset of Figure 2b. The extent of absorption resonance decreases as *u* increases. The maximum absorptance $A = 82$ % is obtained for $u = 1.1$ µm and $d = 350$ nm, whereas the increase of *d* up to 850 nm shifts the absorptance peak ($A = 74$ %) to a higher pitch ($u = 1.4$ µm). It is noteworthy that the peak resonance seems insensitive to the array pitch, due to the sparse nature of NW arrays for *d* below 375 nm. Therefore, light coupling predominantly occurs in the localized radial resonant modes inside NWs rather than as an absorption resonance from the array periodicity. The dominant role of the waveguide modes in the absorption enhancement translates into a strong correlation between *d* and the incident wavelength.

The absorption spectral tunability as a function of *d* for a fixed pitch ($u = 1.1$ µm) is displayed in Figure 2c. Multiple features are observed. First, when *d* increases from 150 nm to 400 nm the absorption peak shifts from 1.0 µm to 2.2 µm (indicated by * in Figure 2c). However, a strong broadening of the peak is observed for larger *d*, eventually leading to a quasi-planar absorption. A detailed evolution of the corresponding mode is presented in Figure 2d where the resonant wavelength scales directly with *d*. Second, for *d* larger than 300 nm a second absorption peak develops in the 0.9-1.2 µm range (indicated by ↓ in Figure 2c) emerging from an extended broadband reaching Ge band gap (solid spheres in Figure 2d). Third, the maximum peak intensity progressively decreases with increasing wavelength, following almost a similar trend as the reference as-grown GeSn layer (black-dashed line in Figure 2c). Fourth, near Ge direct band edge (1.6 µm), the observed asymmetry in the absorptance peak (a sharp change between a dip and a peak) is a signature of Fano-resonance,[39] which dominates the absorption enhancement at longer wavelength (*i.e.* larger *d*). To better understand the radial LMR observed in the absorption map in Figure 2b, a simplified Lorentz-Mie scattering simulation was performed to explore the correlation between the radial modes and the geometry of an infinitely long cylindrical wire. The absorption efficiency of different single GeSn NWs for variable *d* between 100 nm to 600 nm is discussed in Supporting Information (S3). The number of resonant modes increases with *d*, in agreement with the results in Figure. 2c. At *d* between 100 and 200 nm (only 2 LMRs are visible and have the highest absorption efficiency in the 1.55-2.05 μm range, while with a further increase in *d* to 600 nm the main peak shifts to 2.2 μm and the number of LMR increases up to 4 (Supporting Information S3). Coupling among NWs and optical cross-talk[40] promotes the excitation of additional LMR when the NW density increases. The absorption peaks at ~90 % at diameters that are multiple of $\lambda/4$=250 nm, where the light is more efficiently coupled to NWs. For a fixed pitch of 1.1 μm, a decrease in the maximum absorption at the resonance frequency is noticeable in Figure 2c when *d* increases from 250 nm (~ 90 %) to 400 nm (~68 %). Interestingly, when *d* decreases below 300 nm the absorption above 1.6 μm is rapidly suppressed. This indicates that the NW absorption cross section becomes too small to allow an effective absorption in this wavelength range. This is consistent with earlier observations in other material systems,[28,41–45] and occurs when the absorption cross-sections is larger than the NW geometric cross-section.

The effect of the NW height *H* on the absorptance is shown in Figure 2e for *d* = 350 nm and *u* = 1.1 μm. When *H* is lower than 0.2 μm negligible absorption is observed and only wavelengths below 1.6 μm are absorbed in Ge. For comparison, the measured absorptance of the as-grown 1 μm-thick Ge layer is also shown in Figure 2e. By increasing *H* from 0.4 to 2 μm, an absorption peak develops with a gradual increase in intensity up to ~86 %, while the resonant wavelength ($\lambda_{res}$) shifts from 1.7 to 2.0 μm (Figure. 2e). This effect is related to the presence of Fabry-Pérot (FP) modes in the vertical NWcavity. The latter can be highlighted by plotting the absorptance as a function of $H$ (Figure 2f), where local maxima (indicated by black arrows) are separated by ~0.8 μm. Resonant modes are allowed when $H$ is a multiple of $m\lambda/2n$, where $m$ is an integer and $n$ is the effective refractive index. Considering $n$=4.1 for GeSn, at a wavelength of 2.0 μm the FP spacing is ~0.24 μm, in agreement with the local maxima of 0.56 ($m = 2$) and 1.3 μm ($m = 5$) indicated in Figure 2f. Therefore, the proposed NW array design results in a sizable enhancement in the GeSn absorption at 2.0 μm wavelength when $H$ exceeds 0.7 μm.

**Optical properties of GeSn NW arrays.** The simulations outlined above were used to guide the fabrication of GeSn NW arrays and investigate their optical response in the e-SWIR range. As highlighted earlier in Fig. 1, the NWs are tapered and thus the FDTD simulations should also include this morphological feature to allow a more accurate correlation with the experimental observations. Tapering induces a broadening of the LMR peaks ( see Supporting Information S4) while it is expected to minimize the reflectivity[46] and increase light absorption.[47] Additionally, tapering increases the NW fill Factor (FF) from ~2 % at the top to 5 % at the bottom in average for all arrays (see Supporting Information S5). The FF below 5 % suggests that the arrays in our geometrical model are sparse, resulting in very minimal coupling between NWs.[48] Therefore, from the near-field simulations, we conclude that short wavelengths are absorbed in the top part of GeSn NWs, while at longer wavelengths the tapering contributes the strongest to the absorption. The straight cylindrical NW arrays show a well-defined, diameter-dependent absorption resonance, whereas tapered NW arrays show a broad absorption over the entire wavelength range. This behavior agrees well with earlier observations for Si NWs.[49] Tapered NWs consist of a continuous range of diameters along the NW axis; hence, it is possible to absorb almost the entire wavelength range in the e-SWIR. Notably, tapering introduces mode broadening, which translates to a broader resonance peaks in absorptance, when compared to cylindrical NW. However, the magnitude of the absorptance is minimally affected. Data points representing the four fabricated arrays are overlaid (green square for S1, red pentagon for S2, blue star for S3 and purple circle for S4) in the absorption map in Figure 2b to highlight the expected absorption. Measurements on these NW arrays were performed at room-temperature in an integrating sphere optical setup (see Methods for more details).

The experimental absorptance spectra recorded for the four arrays are plotted in Figure 3a-d together with the simulated FDTD absorptance spectra (dashed black-line). In planar GeSn, the absorptance gradually decreases from 40 % at 0.9 μm to less than 5 % at 2.4 μm (grey curve in Figure 3a). The oscillations between 1.6 and 2.1 μm result from FP resonance between Ge and GeSn layers. In NW arrays with smaller diameters (S1 and S2), the absorption is enhanced by a factor of ~2 compared to planar geometry for wavelengths shorter than 1.6 μm, *i.e.* below the Ge band gap. However, a monotonic decrease in the absorption is visible at longer wavelengths (Figure 3a-b). For the arrays with larger diameters S3 and S4, an additional absorption peak develops above 1.6 μm with a maximum intensity at ~2.0 μm, which is more than 1.6 times higher than in planar GeSn (Figure 3c-d). Regardless of the diameter, the measured absorptance is in very good quantitative agreement with FDTD simulations (dashed lines), thus confirming that NW arrays strongly enhance the absorption.

We note that the small offset in S3 and S4 peak positions can be explained by light scattering due to the presence of parasitic NWs, Ge layer over-etching, and the fluctuations in diameter and tapering (see Supporting Information S5). In a sparse NW array (with a FF below 5%), radial LMR are the primary mechanism responsible for resonant absorption.[50,51] Harnessing these resonant modes in GeSn materials can in principle permit the absorption enhancement in the e-SWIR spectral range. Accordingly, to validate that radial LMR is the responsible mechanism for light absorption enhancement, modal numerical simulations were performed to decouple the strongly wavelength-dependent effects of optical energy propagation along NWs, diffraction by the periodic array (*i.e.* photonic crystal effects), and FP-type resonances (reflections at top/bottom sections of the NWs). To achieve high absorption in NWs, the modes must couple to incident plane waves, strongly resonate between the top (air) and bottom (Ge) interfaces, and their absorbed energy needs to be confined within the NW. The resonance modes, indicated by arrows in Figure 3a-d, are calculated for each structure. The peak position of the resonance modes is extracted from the electric field intensity as a function of wavelength in all the NW arrays, as shown in the grey shaded region in Figure 3a-d.

The resonance modes have low-quality factor (Q) in the range of 30-150. The quality factor Q for each one of the fabricated NW arrays is presented in Table 2. Adjacent peaks are observed, for all the arrays, between 0.9-1.3 μm and give rise to enhanced broadband absorption through strong in-coupling between incident light and the modes due to their low-Q. Light with low-Q wavelengths has been shown to easily couple into the mode.[52–54] Above 1.3 μm, the number of resonance modes increases when the NW diameter increases. For instance, S1 and S2 present only one clear peak (with an average Q of 85), located at 1.9 μm as well as some small oscillations induced by FP resonances. The same behavior holds true for the S3 array apart from the appearance of an additional resonance at 2.2 μm with a low-Q factor of 20. Finally, the S4 array is marked with the presence of two well-defined modes at 1.4 and 1.6 μm, with an average Q of 140. This suggests that higher-Q, sub-wavelength resonators may be achievable for larger diameter structures through the introduction of higher order modes, while gaining spectral bandwidth through a lower-Q fundamental mode, which is shifted to longer wavelengths.

A close examination of the corresponding *x-y* cross-sectional distribution of the electric field ($|E|^2$) for the resonance mode at specific resonance wavelength (indicated by arrows in Figure 3a-d), taken at a height of 45 % of the NW height, provides further insights into the LMR absorption behavior and reveals important features (Figure 3e-h). First, different types of resonant modes are excited in NW cavities: leaky resonance and lower-order two-dimensional resonant modes. Second, different types of mode profiles emerge as the NW diameter increases. In the smallest diameter NW array S1, the $HE_{12}$ one-dimensional leaky higher-order mode with two nodes appear (Figure 3e) at shorter wavelength (peak 1,2 and 3), whereas at 1.4 µm (peak 4) the fundamental $HE_{11}$ mode is excited. The latter shows large modal delocalization to the NW surrounding and has no cut-off frequency. A significant spillover into the surrounding air is evident (panel 4 in Figure 3e). A slight increase in the NW diameter (S2) does not have any significant impact on the electric field profile, except a small broadening of the $HE_{11}$ mode (panel 7 of Figure 3f). Note that in S1 and S2 arrays, the absorbed incident energy is localized fully inside the NW below 1.6 µm (peaks 1- 4) for S1 and S2 (peaks 5-8) (Figure 3a and 3b). Furthermore, only the $HE_{11}$ mode is guided into Ge underneath the NW and strongly enhances the absorption above Ge band gap, as the cut-off wavelength of the $HE_{12}$ mode is below 1µm for S1 and S2 arrays. For the S3 array with larger NW diameter (325/450 nm), the $HE_{12}$ mode appears to be more spatially confined inside the NW below 1.61 µm (panels 9 and 10 in Figure 3g), and it starts to broaden to cover a larger coupling area inside the NW above 1.61 µm (panels 11 in Figure 3g) until finally the $HE_{11}$ mode is excited (panel 12 in Figure 3g). With a further increase in the NW diameter and pitch in S4 array, higher coupling efficiency for the $HE_{11}$ mode of 68 % is reached at 2.0 µm (peak 16). Furthermore, the observed large Fano resonance (peak 15) is a signature of absorption enhancement by the Ge layer underneath the NWs.

For a more detailed understanding of the distinct resonance behaviors of the GeSn NW arrays, the distribution of the Poynting vector ($|S|^2$) in the *z–x* plane was simulated at specific resonance wavelengths (0.97 µm for S1, 1.93 µm for S2, 1.76 µm for S3 and 2.10 µm for S3), as shown in Figure 4a. Incident light is guided to the Ge layer in S2 at a wavelength of 1.93 µm above the Ge band edge. For S1, the guided wave attenuates rapidly and does not reach the bottom and is localized in the top of the NWs. For S3 and S4 arrays, the incident energy is concentrated inside the NW above 1.6 µm, which is due to an increase in the absorption cross-section, enhanced by the resonant modes. Next, by analyzing the calculated $|E_x|^2$ and $|E_z|^2$ distributions in the *x–y* plane at different NW heights (0.1, 0.45 and 0.9 µm), the leaky resonance nature of the modes can be better quantified (Figure 4b). The calculated $|E_x|^2$ distribution in the *x–y* plane confirms much stronger leaky mode ($HE_{11}$) resonances in the top part of the S4 NW (8-fold increase in the electric field intensity). Likewise, the$|E_z|^2$ distribution also shows a strong and broadened leaky guided mode ($HE_{12}$) resonance, at the NW top, as shown in Figure 4b. Interestingly, different mode symmetries are degenerated even at single resonance frequency, showing the $EH_{22}$, $TM_{03}$ and $HE_{13}$ modes at the wavelength of 0.94 µm for S4 NW, while at a higher wavelength of 2.10 µm, the S4 NW shows $TM_{21}$, $TM_{01}$ and $TM_{11}$ mode symmetries (Figure S8a, Supporting Information). The $HE_{11}$ mode is concentrated inside and around the NW and thus should not be affected by the order or position of the vertical NW, as long as the nanowires are not too close together. When the NWs are close, the modes couple together and the absorption peaks begin to broaden.

Yee's mesh-based full-vectorial finite-difference mode solver was used within Lumerical® in order to evaluate the modal effective index ($n_{\alpha}$: $\alpha$ is the number of the mode) for all the investigated arrays. In all samples, $n_{\alpha}$ is a complex number where the real part increases and the imaginary part decreases as the mode number increases (see Supporting Information, Figure S8b), thus indicating that these evanescent modes decay exponentially into the NW array. Besides, higher order modes occur at higher frequencies (*i.e.* shorter wavelengths) and have a reduced radiative loss ($q_{rad}$), or put differently they are harder to couple to from free space, which implies smaller total absorption and narrower spectral width.[55] The radiative loss of the fundamental mode is displayed in Supporting Information S8c. The match of the radiative loss of S3 and S4 arrays and the absorption loss create a critical coupling[56] between the incident radiation and GeSn material, which explains the e-SWIR resonance in S3 and S4 arrays.  Table 2 shows the different resonant wavelengths for the both leaky modes $HE_{11}$ and $HE_{12}$, where the corresponding complex modal effective index $n_{\alpha}$ as well as the radiative loss $q_{rad}$ are presented. The increase in the radiative loss for the $HE_{11}$ mode above 1.6 µm for S3 and S4 arrays is a clear indication of the enhanced incident light coupling to NWs at SWIR wavelengths. For comparison, above 1.6 µm, $q_{rad}$ is below 0.03 for S1 and S2 arrays. This giant increase in absorptance coincides well with the enhancement observed in the radiative loss as discussed earlier.

To further investigate the fundamental mode evolution as a function of the wavelength, we now consider the photonic crystals effect originating from the ordered NW array configuration. Note that there is a close relation between the optical eigenmodes we solve for and the modes solved for in the case of a photonics crystal band structure: the modes we solve for at normal incidence correspond to the modes at the Γ point of the photonic crystal. However, since a scattering problem is considered, we must also solve for the evanescent modes at a given wavelength, whereas for the photonic crystal band structure, only the propagating modes with a real-valued propagating constant $k_z$ are solved for. The dispersion relation for the fundamental $HE_{11}$ mode in each array is presented in Figure 5a. As the wavelength increases, each array mode converges to one specific $HE_{11}$ mode. Note that when the wavelength is smaller than the cut-off between a multi-mode and a single mode regime (1.3 μm for S1, 1.4 μm for S2, 1.5 μm for S3, and 1.8 μm for S4), the mode is expected to show considerable coupling between neighboring NWs, meaning that the NW array would support different modes compared to isolated NWs. To couple to an incident plane wave at normal incidence, a Bloch mode must have an even in-plane field distribution. The electric field distribution for S4 in Figure 4b illustrates that the fundamental mode can be coupled to the incident field to yield high absorption at the resonant wavelengths. Interestingly, the extraordinary high mode coupling obtained in S3 and S4 arrays (see Supporting Information, Figure S8b) explains well the strong absorption enhancement measured above 1.6 μm (Figure 3c-d). The real part of the dispersion relation for the unpolarized modes in the S4 sample is shown in Figure 5b, overlaid with the light line (black-dashed line) that delimits the boundary with the region where the leaky modes transition to guided modes occurs. Near 2 μm (peak 16 in Figure 3d), only 2 modes ($HE_{11}$ and $TE_{01}$) contribute to light absorption inside the S4 NW array. The fundamental $HE_{11}$ mode is fairly flat for high values of $k_z$, but become slightly perturbed as it approaches the light line. This effect is clearer for $HE_{21}$ and $TM_{01}$ modes where the distortion of the parallel wavevector $k_z$ is more pronounced closest to the light line. The nature of each leaky waveguide mode is determined from the normalized electric field distribution at 1.49 μm (grey line in Figure 5b) where 3 additional leaky modes ($HE_{21}$, $TM_{01}$ and $EH_{11}$) contribute to the light absorption shown at peak 14 in Figure 3d.

# CONCLUSION

Light absorption in GeSn NW arrays on Si was investigated by combining theoretical and experimental analyses. These systematic studies confirmed the ability to engineer and enhance light absorption in the e-SWIR range using GeSn NW arrays at a Sn content reaching 9 at. %. Leaky-mode resonances (LMR)-induced field enhancements in NWs yield an absorption of ~70% at 2 μm, which translates into a 3-fold increase relative to as-grown GeSn layers. The NW tapering extends the absorption range and enables a broadband absorption with minimized reflection. The spectral tunability of the LMR showed geometry-dependent, specific absorption resonances in agreement with the finite-difference time domain (FDTD) simulations. The latter unraveled that the $HE_{11}$ and $HE_{12}$ optical modes are excited for all NWs. Further investigations of the spectral evolution of the guided mode revealed that modal broadening was the main physical mechanism linked to the absorption enhancement within NWs with larger diameters. The ability to tailor e-SWIR absorption using all-group IV semiconductor platform lays the groundwork to harness a range of the electromagnetic spectrum that has been heretofore mainly accessible using compound semiconductors. These capabilities enabled by GeSn NW arrays would create valuable new opportunities to engineer Si-integrated, scalable photonic and optoelectronic devices operating in the e-SWIR range that benefit from the very low losses and high optical mode confinement of Si waveguides. Future works will focus on harnessing these enhanced optical response to implement and test nanowire array devices for applications including data communication,[57] biomedical sensing, and imaging.

# METHODS

**Epitaxial Growth of GeSn.** GeSn was grown on a 4-inch Si (100) wafer in a low-pressure chemical vapor deposition (CVD) reactor using ultra-pure $H_2$ carrier gas, 10% monogermane ($GeH_4$) and tin-tetrachloride ($SnCl_4$) precursors.[12,20,58] First, a 1.0 µm-thick Ge-VS was grown with a two-temperature step process at 450 and 600 °C followed by a post-growth thermal cyclic annealing (>800 °C). Next, a 1.1 µm-thick GeSn layer with a graded composition from 6.7 at.% to 9.2 at.% was grown at 320 °C for 270 minutes using a Ge/Sn ratio of ~1290 in gas phase.

**NW arrays Fabrication.** Arrays with variable NW diameters and pitch length were patterned with a Raith e-line electron beam lithography (EBL) with a beam energy of 10kV using a negative resist (ma-N 2400). Next, a $Cl_2$-based reactive-ion etching (a Plasmalab 100 ICP-RIE) process was performed to etch the NWs from the GeSn layer. Lastly, the resist was removed using an Oxygen-based plasma process.

**Optical Characterization.** First, UV/visible absorptance spectra were collected at room temperature using a Perkin Elmer Lambda 900 Series UV/VIS/NIR Spectrometer. Spectra were acquired in the double beam mode with baseline correction to eliminate instrument effects. The instrument used a 150 mm integrating sphere with an InGaAs detector at a resolution of 0.5 nm A detailed description of the different measurement configuration is presented in Supporting Information S6. Second, spectroscopic ellipsometry data were collected with a variable angle spectroscopic ellipsometer (VASE, J.A. Woollam, Inc.) at incident angles of 62° to 84° with a 2° step from the sample normal and photon energies in the range of 0.5–1.4 eV (2.5–900 µm wavelength). The illuminated area of the sample at these angles is approximately 3×7 $mm^2$. The experimental tan(Ψ) and cos(Δ) data as well as the different optical models used are shown in Supporting Information S2.

**Theoretical simulations.** Lumerical FDTD ©, a commercial software, was used to perform 3D, full-field electromagnetic simulations of GeSn NWs. Arrays were constructed using a rectangular 3D simulation region, with periodic Bloch boundary conditions applied in the $x$ and $y$ directions and infinite boundary conditions, rendered as perfectly matched layers (PML), in the $z$ direction. All nanowire structures were modeled as GeSn, using the SE-measured optical properties, and were anchored on an infinite silicon substrate with a 1µm Ge-VS layer. All nanowire structures were triangularly packed at a variable fill fraction, with a geometrical configuration extracted from SEM images in Figures 2b and c and were 1.2 µm in height. Furthermore, tapering was introduced in the FDTD simulation to reduce any possible source of discrepancies between the simulated and the microfabricated NW arrays. A detailed description of the simulation conditions is presented in Supporting Information S4.

# FIGURES CAPTIONS

**Figure 1.** (a) Cross-sectional TEM image of the GeSn/Ge-VS/Si multi-layer heterostructure acquired along the [110] zone axis. The scale bar is 1 μm (b) (224) XRD-RSM map showing the graded composition in GeSn from 6.7 at. % to 9.2 at.%.(c) SEM micrograph of the S1-4 GeSn NW arrays (tilting angle 45°). All the scale bars are set to 1 μm (d) Enlarged view of a single S1 NW from (c) with a scale bar of 300 nm.

**Figure 2.** (a) A 3D scheme of the cylindrical NWs in a triangular lattice. The relevant geometrical properties the height $H$, the diameter $d$ and the pitch $u$ are indicated. The grey shaded area shows the PML regions considered for the 3D-FDTD simulation. The simulated absorptance of the structure presented in (a) for diameter and pitch that varies between 0 and 2.2 μm. The inset indicates the absorptance as a function of the pitch for two fixed diameters (350 dashed-red line and 850 nm, dashed-blue line). (c) The spectral absorptance for a 1.2μm fixed height and a 1.1 μm pitch and different diameters (between 150 and 450nm). The peaks indicated by * and ↓ are the first two excited LMR of the cylindrical NW array. The dashed black line represents the simulated absorptance of the GeSn planar layer. (d) The evolution of the two LMR observed in panel as a function the diameter of the cylindrical NW. (e) The simulated spectral absorptance for a fixed diameter of 350 nm and a fixed pitch of 1.1 μm as a function of the height. The dashed-grey line is the simulated absorptance of only the Ge-VS layer (in other words, when $H = 0$ μm). (f) The absorptance vs the height of the structure considered at panel (e) at a fixed wavelength of 2 μm. The arrows indicate the Fabry-Pérot resonances.

**Figure 3.** (a-d) The spectral absorptance of the 4 NW arrays structures (S1-4) measured with the integrating sphere between 0.9 and 2.4 µm. The corresponding dashed-lines are the simulated FDTD response of the NW arrays, calculated based upon the structure shown in Figure 2a. The numbered arrows correspond to the $HE_{11}$ and $HE_{12}$ resonance mods observed for each structure, as extracted from the electric field intensity as a function of wavelength shown in the grey shaded region for each NW array. (e-f) The corresponding normalized-electric field $|E|^2$ distribution for each corresponding resonance peak of the $HE_{11}$ and $HE_{12}$ modes. The panel numbers are correlated to the resonance peak position in panel (a-d) and are presented in Table 2. The scale bar for all the electric field maps is set to 100 nm.

**Figure 4.** (a) Calculated Poynting vector distribution in the $z - x$ plane at resonance wavelengths of 0.97, 1.93, 1.76 and 2.10 µm for the tapered NW array structures S1-4, respectively. The color scale for background profile indicates the relative magnitude of Poynting vector, normalized by the maximum magnitude. The scale bar is equal to 100 nm. (b) Calculated $|E_x|^2$ and $|E_z|^2$ distribution in the *x–y* plane of the S4 NW array at different *z* position, indicated in the panel (a) S4 by the horizontal dashed lines. Three different heights $z$ (0.1, 0.45 and 0.9 µm) are analyzed. The $|E_x|^2$ field patterns correspond to $HE_{11}$ mode, whereas the $|E_z|^2$ correspond to $HE_{12}$ mode. the scale bar is 100 nm.

**Figure 5.** (a) Dispersion of the optical fundamental mode $HE_{11}$ in all the NW arrays (S1-S4). (b) Dispersion relation for the guided eigenmodes of the S4 NW array with a diameter of 375/550 nm. (c) the corresponding optical modes excited by the S4 NW array at a wavelength of 1.49 µm (the grey line in panel (b)). The scale bar is 100 nm.

## ASSOCIATED CONTENT

### Supporting Information.

The Supporting Information will be available free of charge on the ACS Publications website.

Additional information including experimental and theoretical details, additional figures, and additional references.

## AUTHOR INFORMATION

Corresponding Authors:

*E-mail: anis.attiaoui@polymtl.ca ; oussama.moutanabbir@polymtl.ca

Notes

The authors declare no competing financial interest.

## ACKNOWLEDGEMENTS

The authors thanks J. Bouchard for the technical support with the CVD system, M. Attala for fruitful discussions, B. Baloukas for support with the integrating sphere measurement. O.M. acknowledges support from NSERC Canada (Discovery, SPG, and CRD Grants), Canada Research Chairs, Canada Foundation for Innovation, Mitacs, PRIMA Québec, and Defence Canada (Innovation for Defence Excellence and Security, IDEaS). S.A. acknowledges support from Fonds de recherche du Québec-Nature et technologies (FRQNT, PBEEE scholarship).

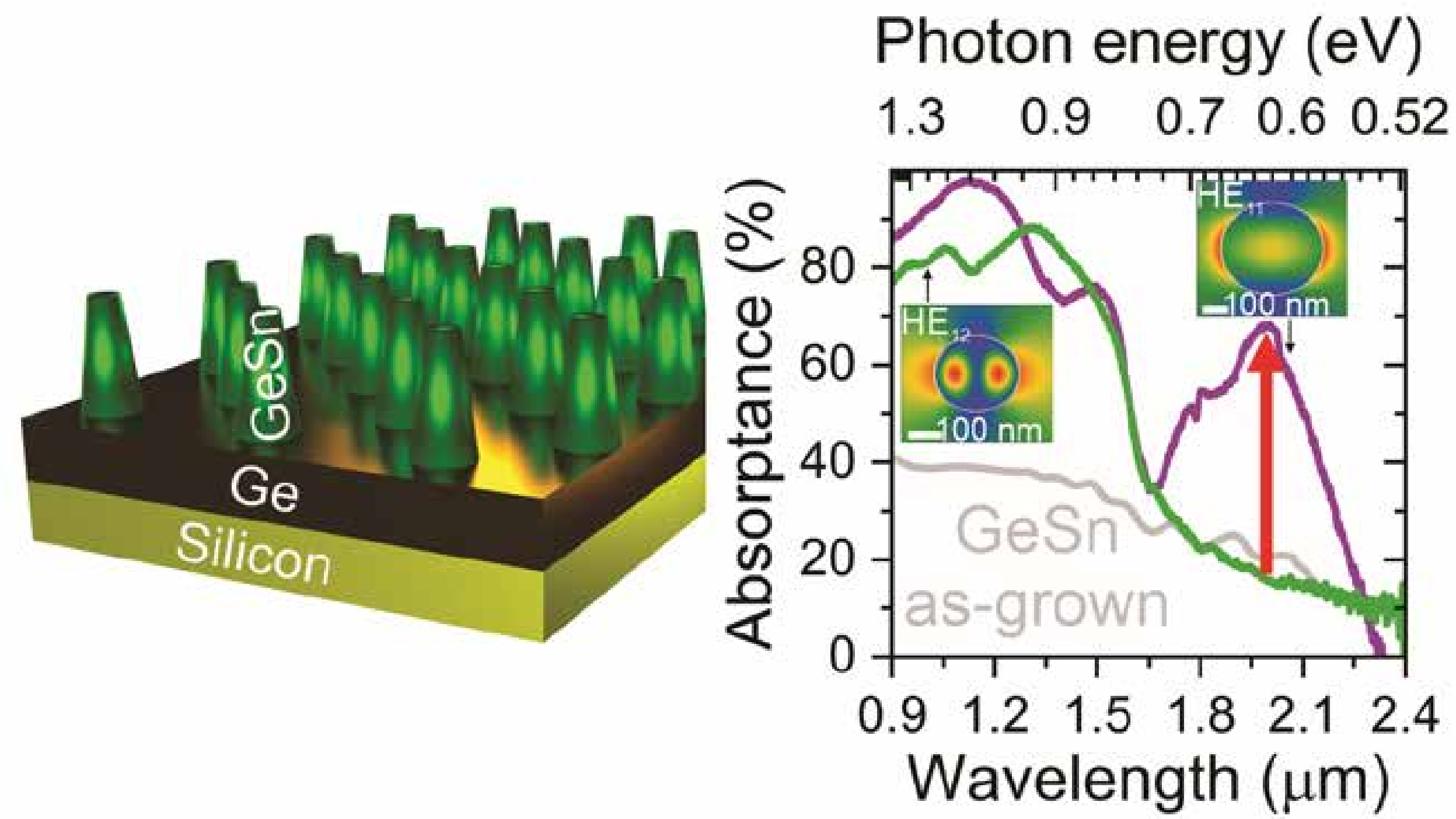

TOC Graph

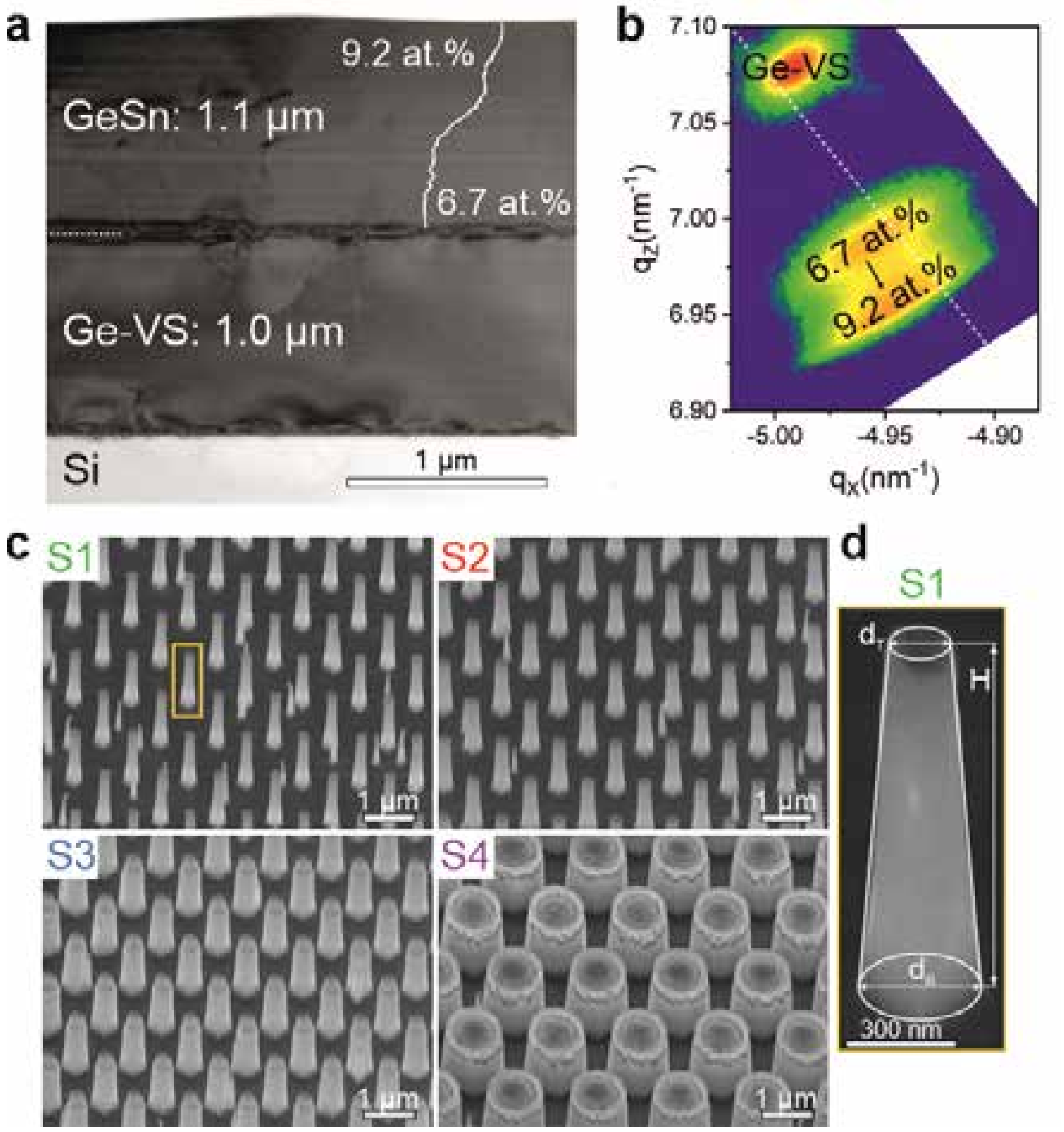

Figure 1

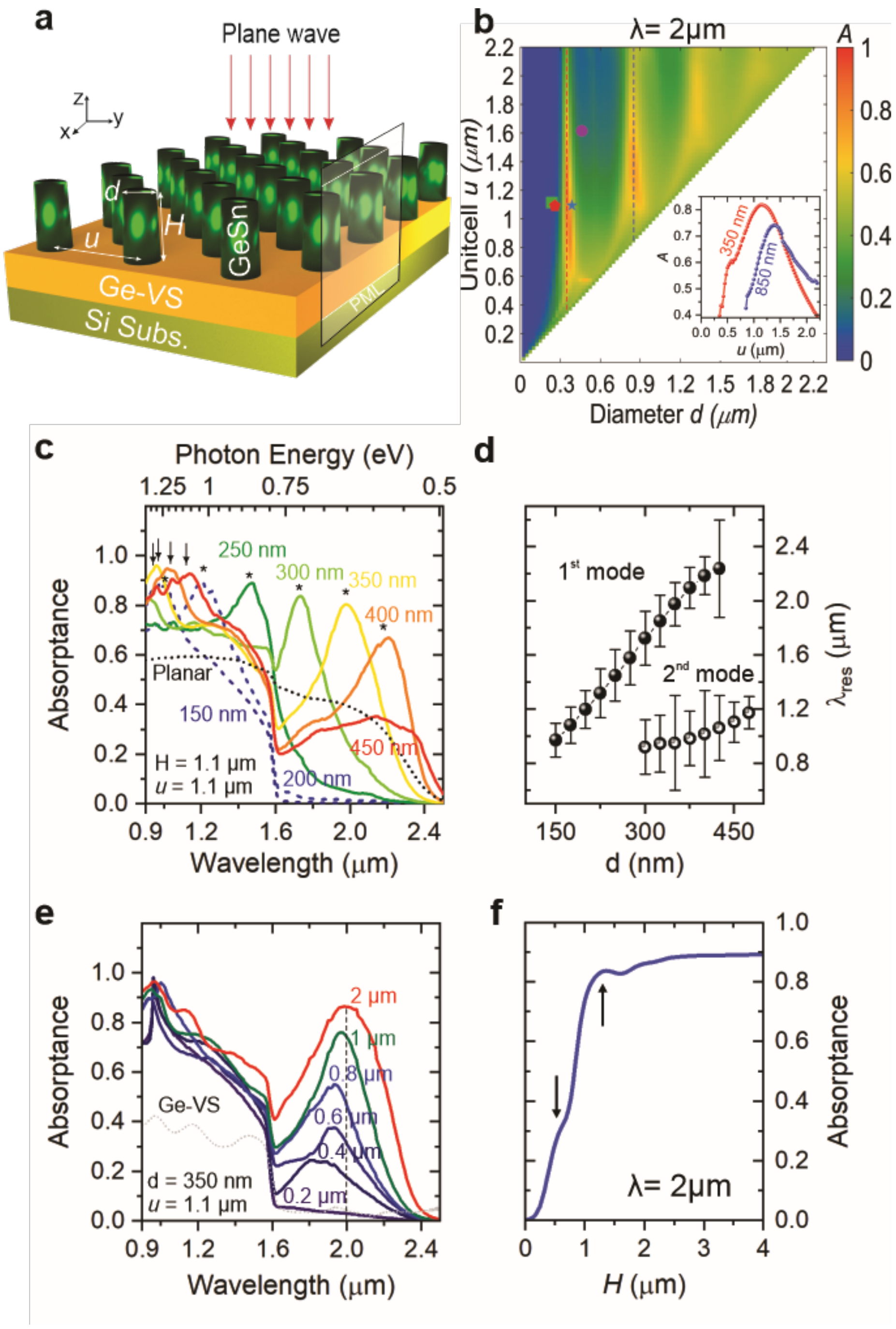

Figure 2

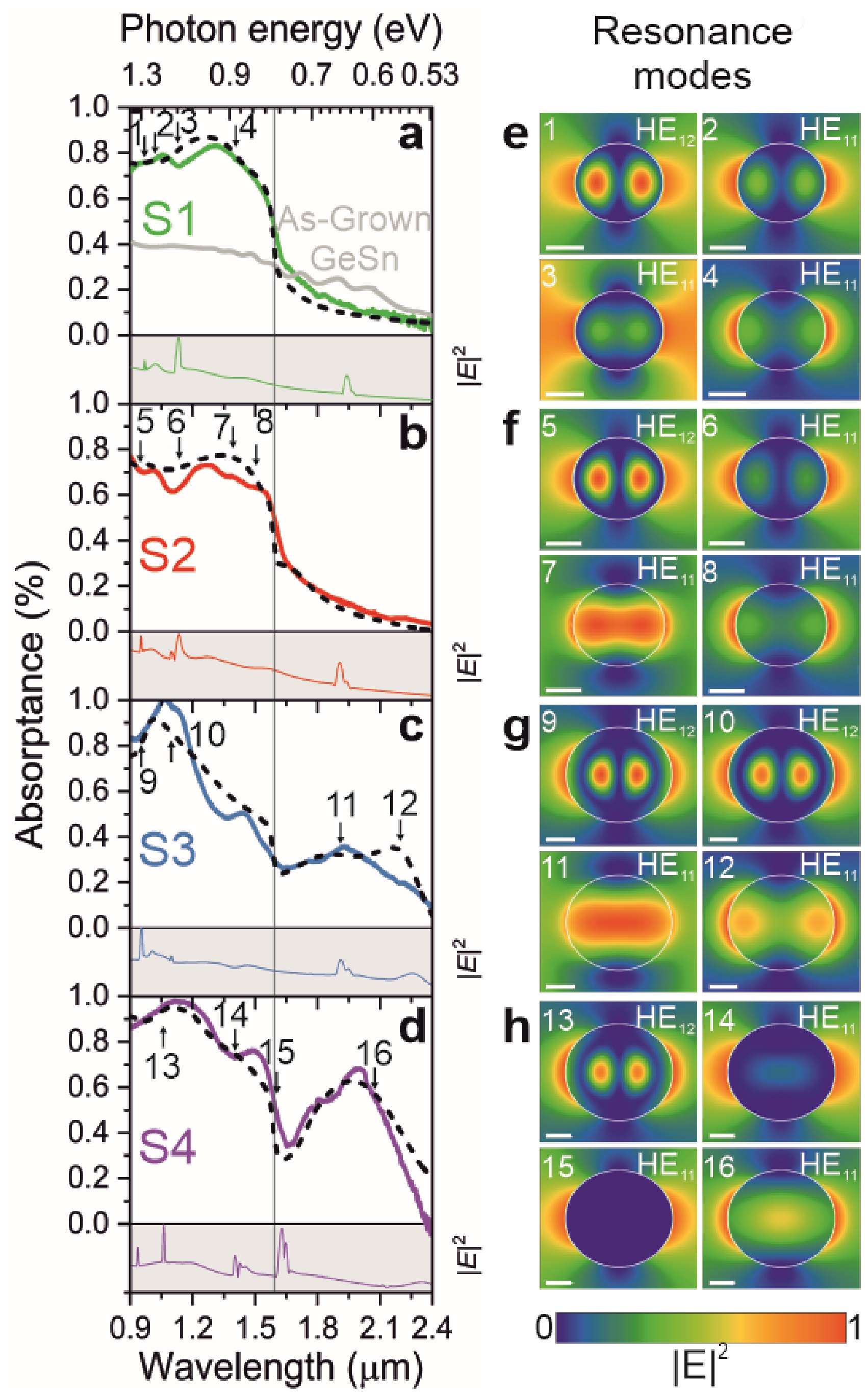

Figure 3

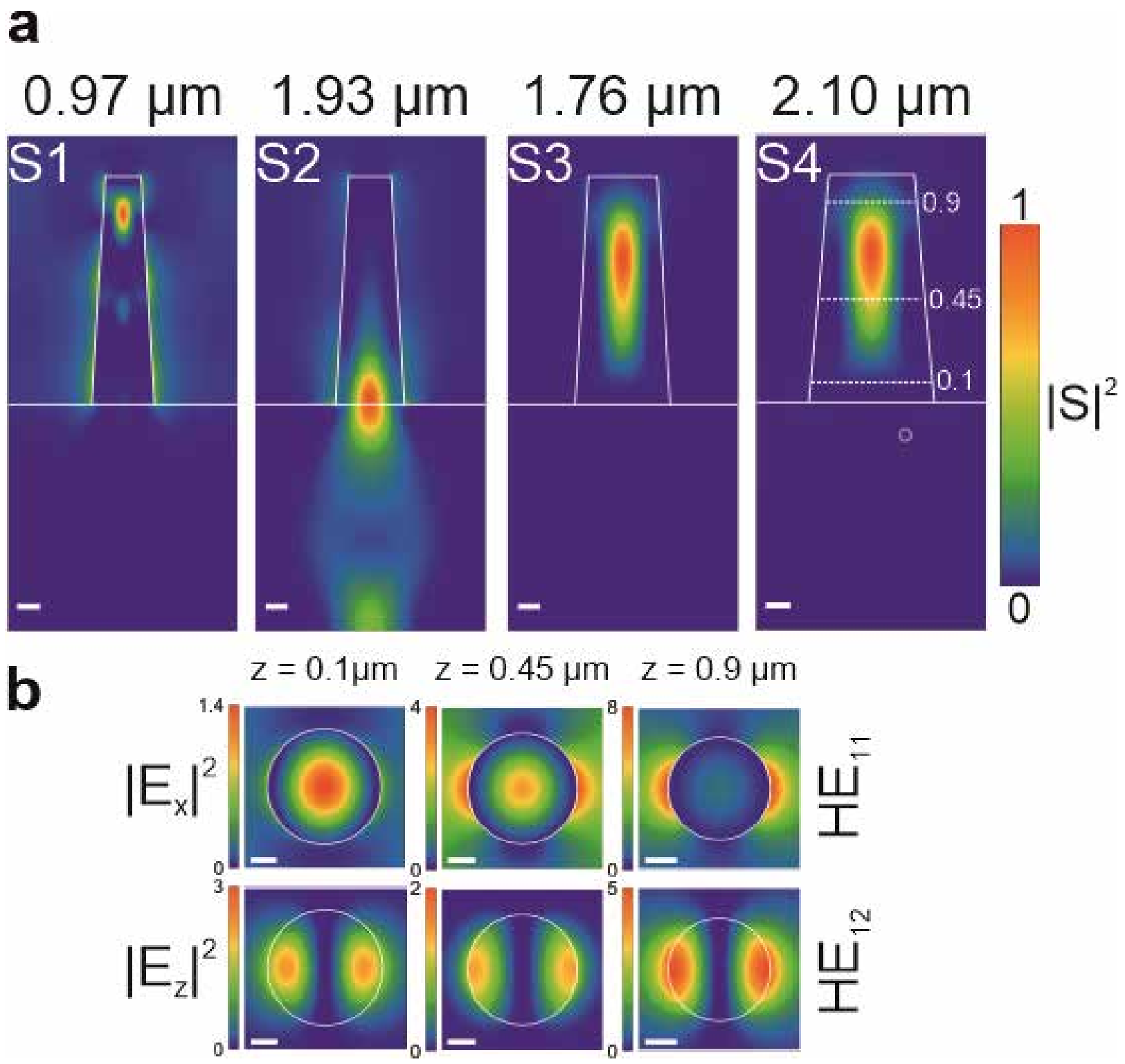

Figure 4

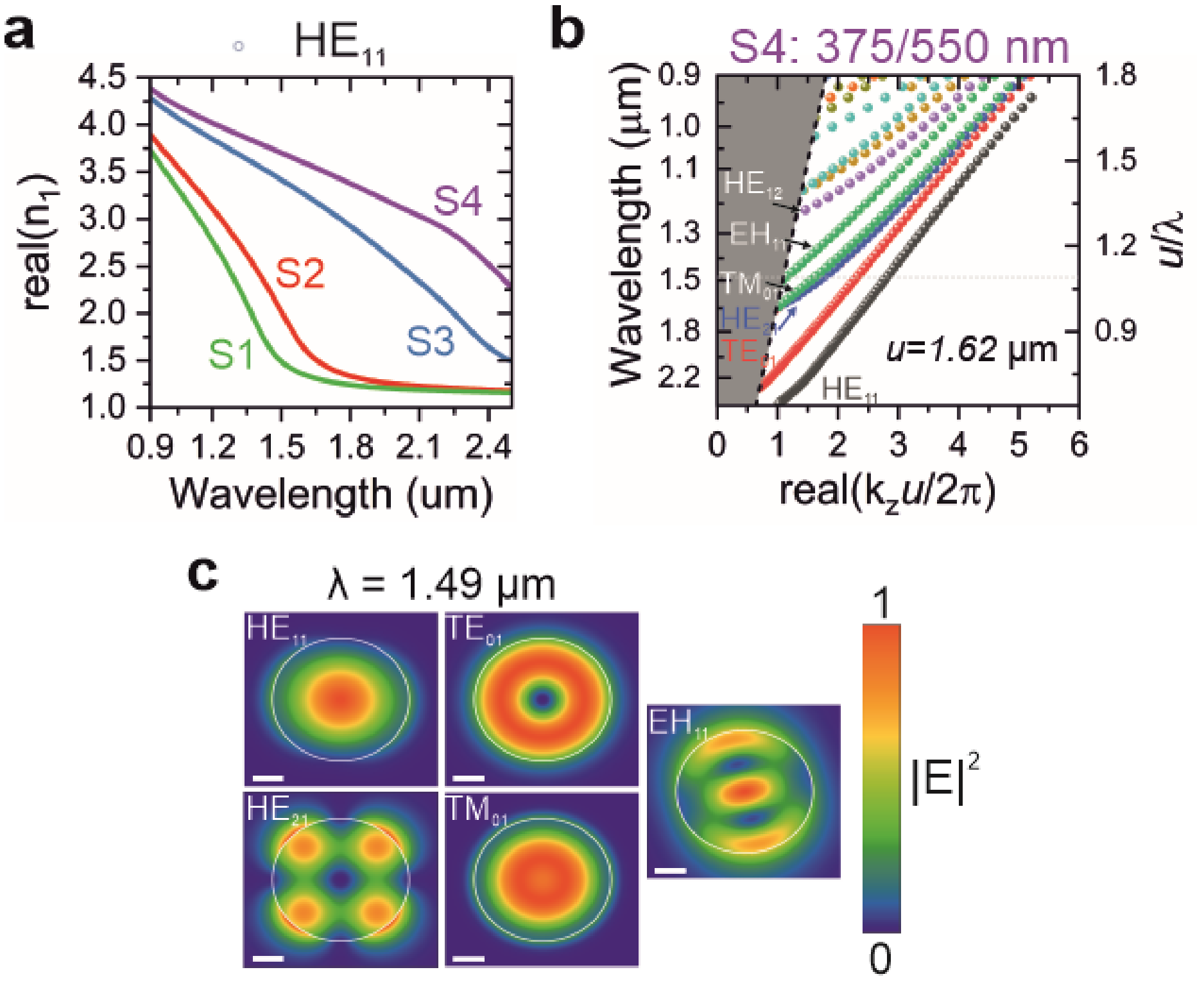

Figure 5

Table 1: Structural information of the different fabricated NW arrays extracted from the statistical SEM images analysis from Figure 1c.

| params / NWs | GeSn NW Arrays | | | | |
|---|---|---|---|---|---|
| | $d_T( nm)$ | $d_B( nm)$ | $u\ (\mu m)$ | H (μm) | FF (%) |
| S1 | 175 | 300 | 1.1 | 1.2 | 2.1 |
| S2 | 200 | 325 | 1.1 | 1.2 | 2.6 |
| S3 | 325 | 450 | 1.1 | 1.2 | 5.6 |
| S4 | 375 | 550 | 1.6 | 1.2 | 3.7 |

Table 2: Fundamental $HE_{11}$ and $HE_{12}$ excited mode resonances for all the GeSn NW nanostructures (S1-4).

| params / Array | GeSn NW Arrays | | | | | |
|---|---|---|---|---|---|---|
| | Modes | Peak # | $\lambda_0 (nm)$ | $n_\alpha$ | $q_{rad}$ | $Q$ |
| S1 | $HE_{12}$ | 1 | 970 | 1.15-i1.66 | 1.445 | 143 |
| | $HE_{11}$ | 2 | 1021 | 3.42-i0.33 | 0.096 | 30 |
| | $HE_{11}$ | 3 | 1128 | 3.03-i0.32 | 0.107 | 45 |
| | $HE_{11}$ | 4 | 1409 | 1.79-i0.28 | 0.160 | N.A. |
| S2 | $HE_{12}$ | 5 | 950 | 2.03-i0.69 | 0.340 | 88 |
| | $HE_{11}$ | 6 | 1130 | 3.32-i0.30 | 0.090 | 40 |
| | $HE_{11}$ | 7 | 1250 | 2.95-i0.29 | 0.100 | N.A. |
| | $HE_{11}$ | 8 | 1460 | 2.18-i0.30 | 0.138 | N.A. |
| S3 | $HE_{12}$ | 9 | 1005 | 3.12-i0.35 | 0.112 | 20 |
| | $HE_{12}$ | 10 | 1099 | 2.92-i0.35 | 0.120 | 70 |
| | $HE_{11}$ | 11 | 1946 | 2.63-i0.22 | 0.084 | 50 |
| | $HE_{11}$ | 12 | 2202 | 2.11-i0.16 | 0.076 | 10 |
| S4 | $HE_{12}$ | 13 | 1060 | 3.64-i0.30 | 0.083 | 65 |
| | $HE_{11}$ | 14 | 1405 | 3.80-i0.21 | 0.055 | 74 |
| | $HE_{11}$ | 15 | 1627 | 3.56-i0.20 | 0.056 | 113 |
| | $HE_{11}$ | 16 | 2113 | 3.02-i0.18 | 0.059 | 10 |

# Supplemental Material:

# Tunable Extended-SWIR Absorption in GeSn Nanowire Arrays

A. Attiaoui,[1,*] É. Bouthillier,[1] G. Daligou,[1] A. Kumar,[1] S. Assali,[1] and O. Moutanabbir[1,*]

[1] Department of Engineering Physics, École Polytechnique de Montréal, C. P. 6079, Succ. Centre-Ville, Montréal, Québec H3C 3A7, Canada

## Contents

## S1. Effect of NW Etching on Lattice Strain.

In the XRD 2θ-ω scan around the (004) order (Figure. S1a), the Ge-VS peak is visible at 66.06°, while that of GeSn is detected at angles lower than 65.5 °. In the measurements performed on the S1:175/300 nm NW array, a shift of ~0.1° to larger angles is observed for the 9.2 at.% GeSn peak region, which indicates that the NW etching promotes additional strain relaxation in GeSn. Similarly, the presence of a Ge shoulder peak at ~66.0 ° indicates that the prolongated-$Cl_2$ etch reached the Ge-VS, thus inducing full-strain relaxation in part of the layer from the original (tensile) of $+0.16$ %. The HR-RSM map in Figure S1c confirms this relaxation of Ge-VS upon etching. We note that due to the reduced GeSn volume after NW etching its signal falls below the detection limit of the XRD setup to allow clear RSM measurements. The observed relaxation and Ge over-etching justified the modification of the FDTD model (Figure S1c) to include an additional Ge segment of a thickness $d_{Ge}$ in the bottom section of the NW.

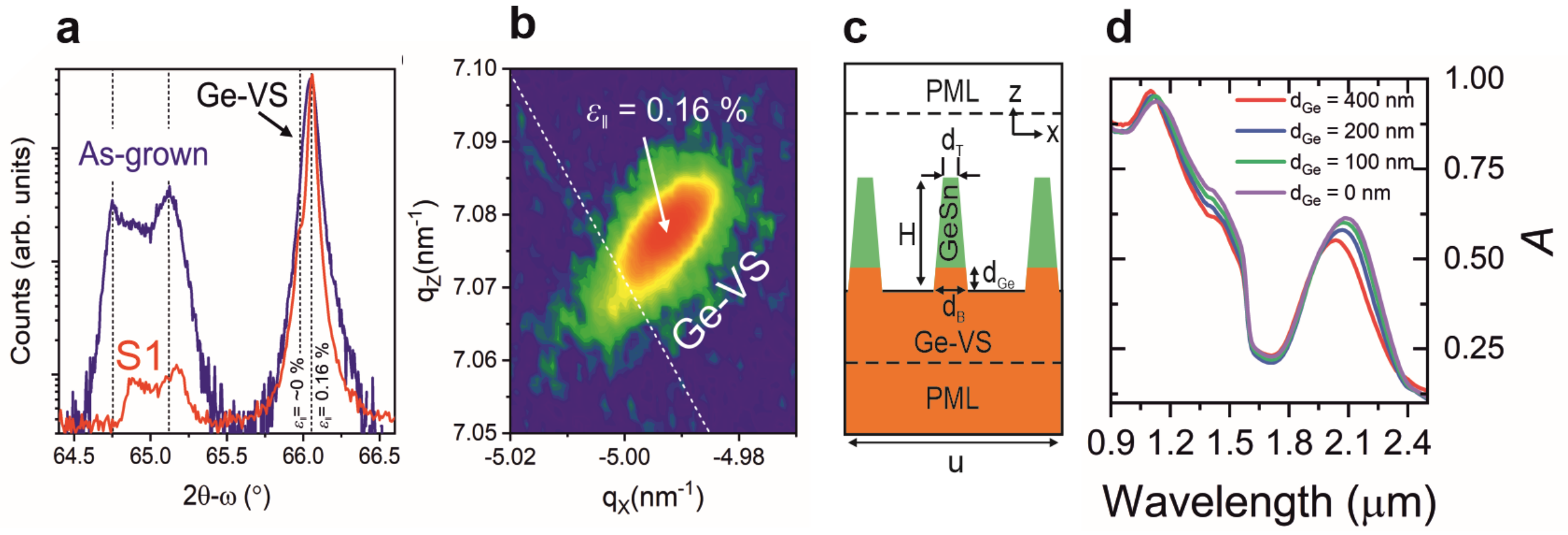

**Figure S1. (a)** XRD 2θ-ω scans around the (004) order for the as-grown GeSn layer (blue curve) and the S1 (175/300 nm) NW array. The strain relaxation induced by the NW etching induces the 0.1° shift to larger angle in the 9.2% peak. Second, the Ge layer over-etching induces full-strain relaxation in part of the layer. **(b)** The corresponding RSM map around the asymmetrical (224) reflection where the Ge layer is 0.16% tensile strained. **(c)** FDTD model including the Ge segment in GeSn NWs. **(d)** FDTD absorptance spectra where $d_{Ge}$ is varied from 0 to 400 nm for a height of 1.3 μm highlighting the effect of the Ge over-etching.

Interestingly, from Figure S1d, increasing the over-etched Ge thickness from 0 to 400 nm induces a decrease in the absorptance at 2μm and an increase near the resonance mode close to 1.1 μm. The over-etching is also accompanied with a small blue shift for the resonance mode near 2 μm.

## S2. Optimization of the Ellipsometric Optical Model.

In the RSM map (Fig.1b) compositional grading from 6.8 to 9.2 at. % Sn is visible, which is further highlighted in the EDS profile displayed below in Figure S2a. Within the 1.1 μm-thick GeSn layer two regions with different Sn content can be identified with a thickness of 623 nm (6.8 at. % Sn) and 455 nm (9.2 at. % Sn). The graded composition was then incorporated in the spectroscopic ellipsometric (SE) optical model. The variable angle spectroscopic ellipsometry (VASE) uses change in the state of polarization of light upon reflection for characterization of surfaces, interfaces, and thin films. In ellipsometry, the measured ratio ρ of the reflection coefficient $r_p$ and $r_s$ can be expressed in terms of the amplitude ratio tan Ψ and the phase angle Δ:[1]

$$\rho = \frac{r_p}{r_s} = \tan\Psi\, e^{-i\Delta} \qquad (1)$$

where the two ellipsometry parameters $\Psi$ and $\Delta$ can be obtained directly from the SE measurements; $r_p$ and $r_s$ represent parallel and perpendicular reflection coefficients to the plane of incidence, respectively. Different SE optical models were considered based on the structural characterization. Graded layers are simulated in the WVASE model by breaking the layer into n+1 sublayers. The thickness of the $i^{th}$ sublayer is defined as:

$$d_i = \begin{cases} \frac{D}{2n} & ; i = 0, n \\ \frac{D}{n}; & i = 1 \ldots n-1 \end{cases}$$

where D is the GeSn total layer thickness. For the linearly graded layer, the Sn content $x_i$ in the $i^{th}$ sublayer will be given by $x_i = \left((n-i)x_0 + ix_n\right)/n$ where $x_0$ and $x_n$ are the Sn content in the #1/#2 interfaces. Furthermore, due to the small gradient, this approach can be successful to estimate the effective dielectric function of the layer. The surface root mean-squared (RMS) roughness of 10.5±1.2 nm, measured on a 20×20 $\mu m^2$ atomic force microscope (AFM) image (Figure S2c), was used as input for the SE optical model to account for the RMS layer shown in Figure S2b. Figure. S2d shows only the raw spectroscopic parameters (Ψ) for each optical model, composed of a Ge-VS/GeSn layer #1/GeSn layer #2 as described in the schematic model in Figure S2b. Incorporating the effect of compositional grading in the layer reduces the MSE from 4 to 0.05, as show in in Figure S2d, which is a clear indication that the optical model encompasses the physical nature of the studied material. Figure. S2b is a schematic of the different optical models considered to extract the optical constant of the GeSn layer. Three optical models were considered with an increasing level of complexity for the $Ge_{1-x}Sn_x$ sample. The first model ($M_1$) neglects the observed Sn gradient by assuming an average uniform Sn composition of 8.5 at.% with the 1.085µm thickness. The second model ($M_2^{gr}$) incorporates the Sn non-uniformity by grading the energy gap as a function of the thickness of the layer in the considered Johs-Herzinger optical model.[2] The third model ($M_3$) separates the single GeSn layers into 2 distinct layers with 6.3 at. % and 9.2 at.%, respectively, and the corresponding thicknesses extracted from EDS as shown in Table S1. Table S1 provides the compressive strain, the thicknesses obtained from XTEM and EDS, and the Sn at. % concentration in each layer of the stack. It is clear that $M_1$ have the highest mean-square error (MSE) of 3.954 which is an indication of its inadequacy. $M_2^{gr}$ shows a staggering 100-fold improvement in the MSE. This can be explained by the increase of the number of fitting parameters in the parametric oscillator model (from 6 in $M_1$, 12 in $M_2^{gr}$ and $M_3$). Furthermore, this approach can be successful only if the gradient is small, which is the case in the current material as shown in the EDS spectra.

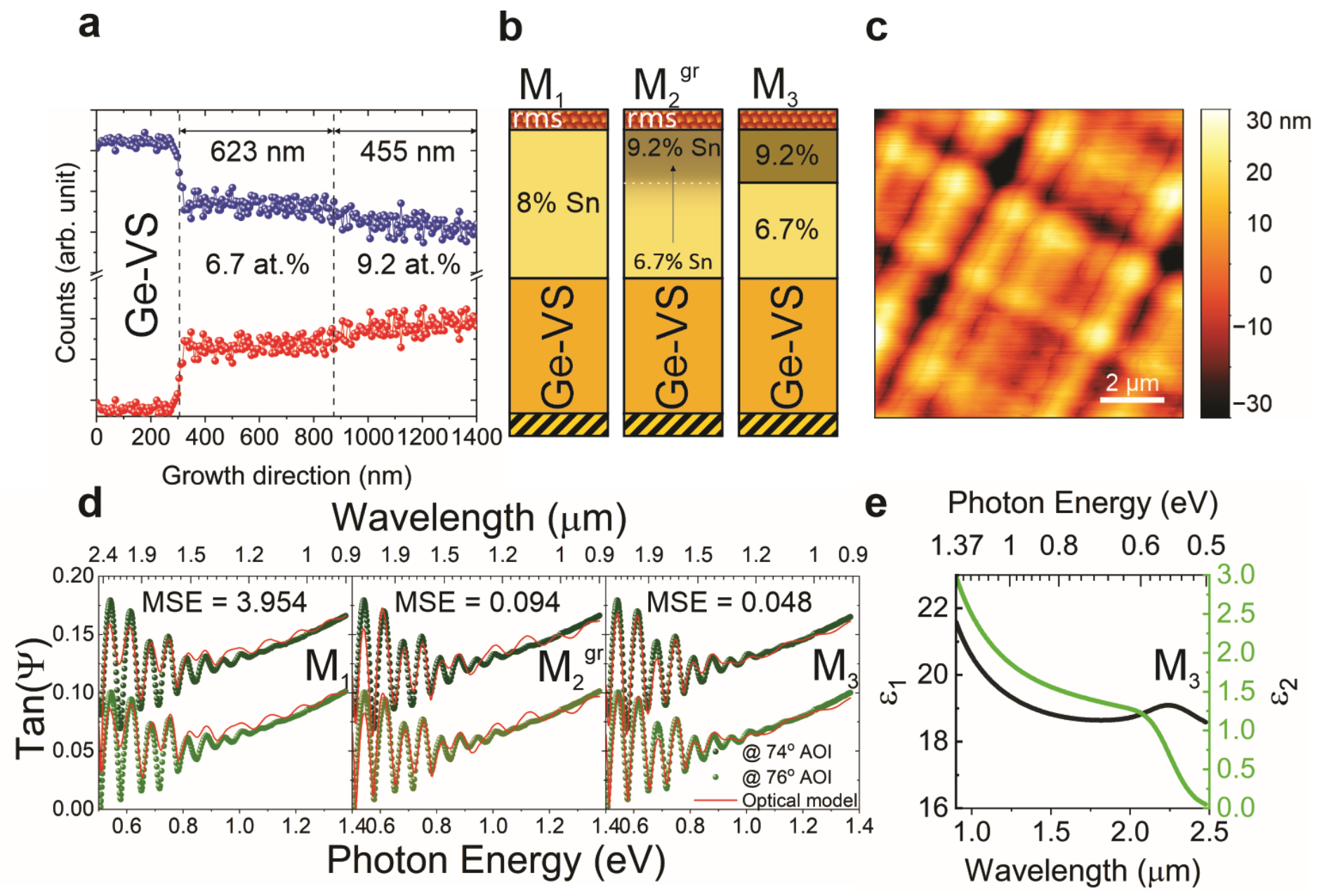

**Figure S2. (a)** EDS profile of the as-grown 1.1 µm-thick GeSn layer on top of the Ge-VS that indicates the presence of a 6.7-9.2 at.% compositional grading. **(b)** Schematic illustrations of the different optical models $M_1$, $M_2^{gr}$ and $M_3$ considered to incorporate the effect of Sn composition grading in the dielectric function measured with SE in the spectral range of 0.9 to 2.5 µm. (c) A 20×20 µm$^2$ atomic force microscope (AFM) map with RMS=10±1 nm. **(d)** The measured (in green) and modelled (in red) SE parameters (tan Ψ) for the 3 different models presented in panel **(b)** as well as the MSE. **(e)** the dielectric function of the as-grown GeSn layer extracted from the $M_3$ model between 0.9 and 2.5 µm.

If the gradient is high, a large number of very thin layers would be required in order to calculate the effective dielectric function resulting in severe numerical complications.[3]

| params / layer | Ge − VS/GeSn =M00715 | | | |
|---|---|---|---|---|
| | Sn (%) | $\varepsilon_{\parallel}$ (%) | $d$ (nm) XTEM | $d$ (nm) EDS |
| Ge-VS | N.A. | 0.147 | 1065.25 | N.A. |
| #1 | 6.7 | -0.182 | 1100.85 | 622.9±1.3 |
| #2 | 9.2 | -0.350 | | 455.3±1.9 |
| RMS | 10±1 nm | | | |

Table S1: Structural information of the GeSn layer: Sn at.% and epitaxial strain were extracted from HRXRD-RSM map shown in Fig. 1b analysis, cross-sectional TEM (XTEM) thickness of the complete layer as well as the individual graded layer from EDS.

## S3. Lorentz-Mie Scattering Formalism.

To quantify the absorption and scattering efficiencies in a single cylindrical nanowire configuration, the corresponding absorption and scattering expansion coefficients were evaluated based on the Lorentz-Mie scattering formalism as detailed in Ref. **4**. To this end, the electric $\vec{E}$ and magnetic fields $\vec{H}$ inside the NW, in addition to the incident and scattering fields were calculated. Figure S3 highlights the unpolarized absorption efficiency $Q_{abs}$ of an infinitively long GeSn single NW for different diameters $d$ that varies from 100 nm to 600 nm with a 20 nm step. LMR are clearly observed and occurs at specific wavelengths as expected. Increasing the density of the NW (by increasing the FF) in the array would induce coupling between the NWs. However, this effect cannot be quantified using a simple Mie-Lorentz scattering formalism and it would require a more elaborate 3D simulation that is out of the scope of the current investigation. Thus, the enhanced light coupling with NWs compensates the reduction in the extinction coefficient $k$ at longer wavelengths and allows for a tunable broadband absorption spectrum. Notably, the absorption is more than double that of the planar 1.1 µm $Ge_{0.91}Sn_{0.09}$ thin film above the Ge-VS direct band gap of 1.6 µm (black dashed-line in Figure 2c)

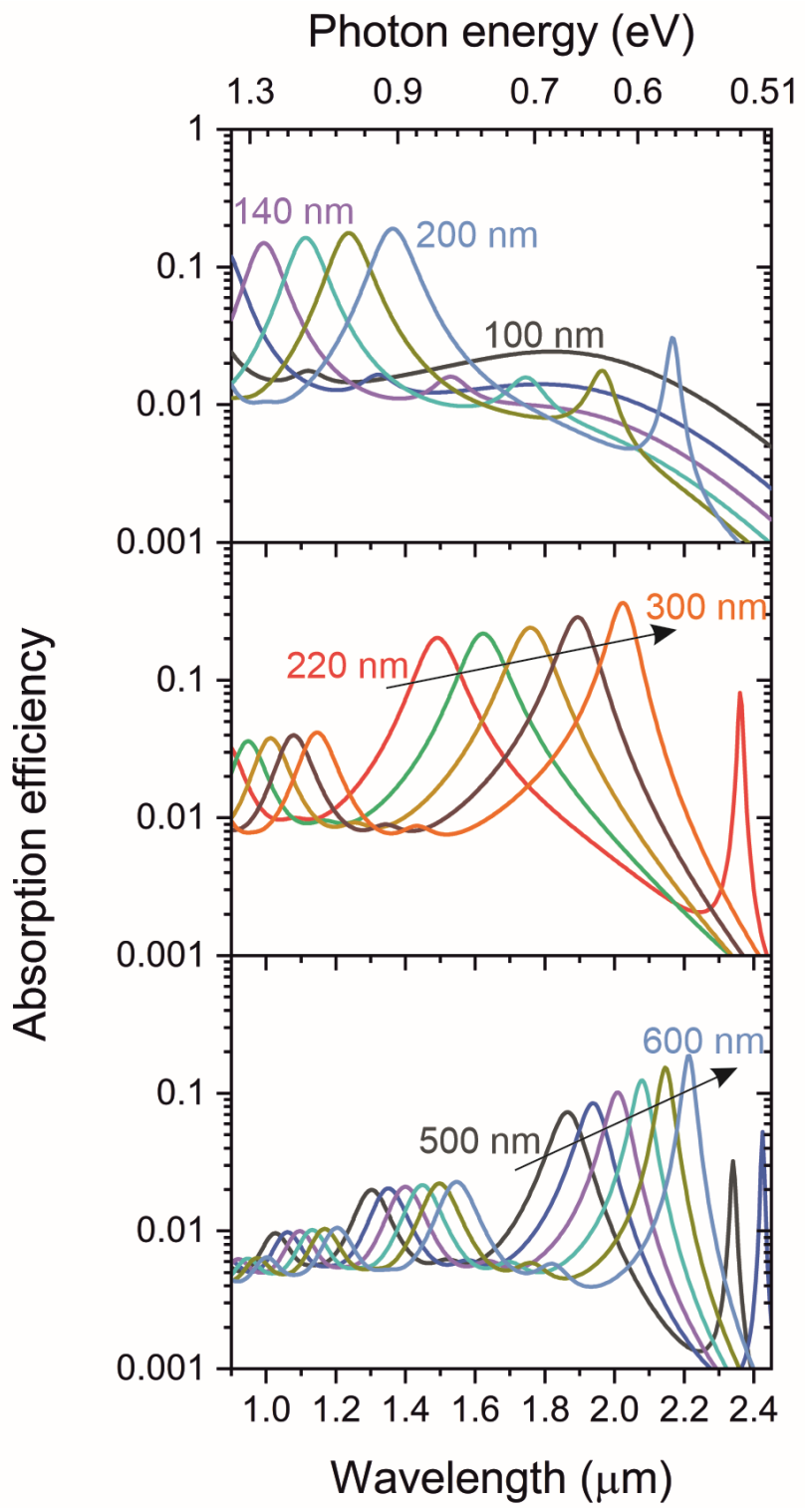

**Figure S3.** Spectral absorption efficiency of an infinitively long GeSn NW, calculated with the Mie-Lorentz scattering formalism, illuminated from the top for different diameters. The diameters are varied from 100 nm to 600 nm with a 20 nm step. The increasing number of LMR is clear when increasing the diameter.

## S4. FDTD Simulations of NW Arrays.

Since the simulations of nanostructured absorbers are focused on the absorption within the active material and not the far-field of the scattering from the structure, perfectly matched layers (PML) were used. A PML is not a boundary condition by itself, it is rather an artificial dielectric [5] with a gradual variation of its properties leading to high attenuation and low reflection. The computational domain consisting of a NW on a 1.1μm Ge-VS was chosen. The domain has a quasiperiodic boundary conditions in the horizontal directions (x, y) and absorbing perfectly matched layers (PML) boundaries in the vertical z- directions, as shown in Figure S4a. The dispersion of the dielectric function of $Ge_{0.91}Sn_{0.09}$ and the Ge-VS layer were extracted from SE measurement as highlighted in section S2 by fitting the refractive index of both semiconductors with a Lorentz model above and below the band gap. The spectral response of the nanowire layer is obtained by using a short 13 fs incident pulse with constant transverse wavenumber and Fourier transforming the simulated time-domain data. The absorptance is computed from the fields in a plane 100 nm above the nanowire and 50 nm in the substrate, respectively.

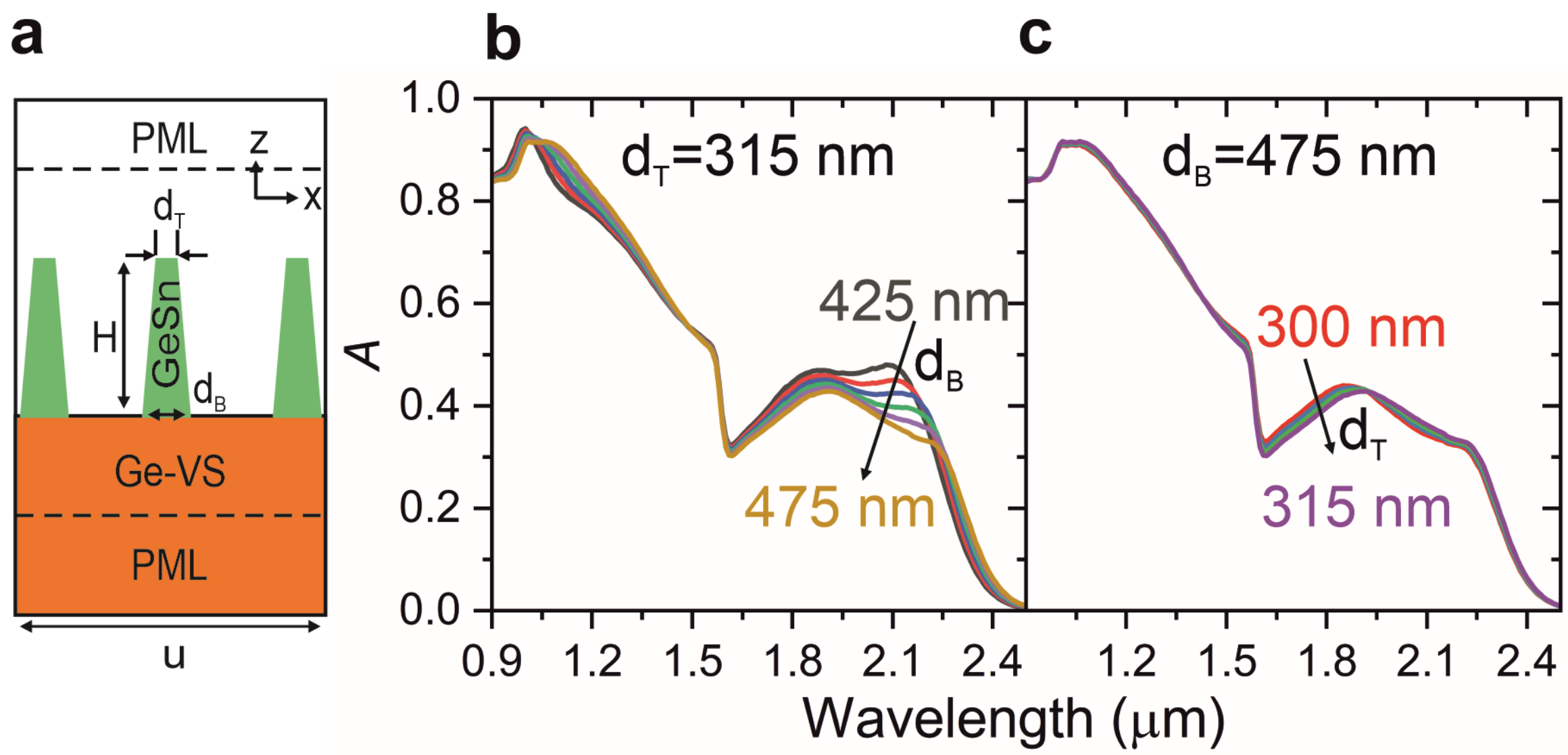

**Figure S4.** **(a)** A simplified FDTD model that shows the relevant part of the simulation, mainly, the PML region and the tapered NWs geometrical parameters. The effect of the geometrical parameters of the tapered NW on the absorptance by varying: **(b)** the bottom diameter ($d_B$) from 425 to 475 nm for a fixed top diameter ($d_T$) of 315 nm, and **(c)** the top diameter ($d_T$) from 300 to 315 nm for a fixed bottom diameter ($d_B$) of 475 nm.

To investigate more in-depth the effect of tapering on the absorptance ($A$), $A$ was evaluated for different geometries where the top and bottom diameters of the NW were varied systematically. In Figure S4b, the top diameter was fixed to 315 nm and the bottom diameter was changed from 425 to 475 nm. The resonance mode below 1.2 μm redshifts and broadens when increasing the bottom diameter whereas the change is negligible when the top diameter is varied. Next, above the Ge band edge (1.6 eV), the effect of increasing the bottom diameter induces a clear broadening of the modal frequency coupled with a reduction of the absorptance. In contrast, increasing the top diameter will reduce slightly the absorptance without any additional resonant modes, as shown in Figure S4c.

## S5. NW Array Geometry and Dimensions.

The volume filling factor or fill-fraction (FF) for a nanocone is given as the ratio between the volume of the nanocone to the volume of the triangular lattice. Thus, FF is equal to

$$FF = \frac{\pi(d_T^2 + d_T d_B + d_B^2)}{12\sqrt{3}u^2} \qquad (2)$$

It is important to highlight the FF for the fabricated arrays is below 5%, as shown in Table 1, which suggests that the NW arrays are sparse, resulting in a very minimal coupling between the NWs. Figure S3a is a digital image of the microfabricated NW arrays (S1-4). The visible redshift is induced by light scattering due to the different NW diameters.

A digital image of the microfabricated NW arrays is displayed in Figure S5a. The top-down fabrication process induced a NW diameter fluctuation as well as tapering within a single array. Therefore, in order to capture the optical response of the array, multi-array geometries were simulated. To accurately extract the geometrical shape of the NW arrays, statistical analysis of multiple SEM images allowed for the estimation of an upper and lower bounds for the top and bottom diameters of the tapered NWs. Figure S5b presents the statistical analysis where the diameter in the middle region of the wire was measured for 20×20 $\mu m^2$ SEM maps. Then, a histogram is fitted with a normal distribution (red curve in Figure S5b) in order to obtain the mean $\mu$ and the standard deviation $\sigma$ of the measured diameters, which can be viewed as a representative estimation of the top and bottom part of the wire. The obtained diameters are shown in Figure S5b. It is important to mention the high standard deviation of 89 nm of the sample S4 which is an indication of non-uniformity in the center of the wire. The chosen value of the top and bottom diameters of the wire lie within 3σ of the mean of the normal distribution. Additionally, higher magnification SEM images (such as the one shown in Figure 3a in the main text) were used to extract more accurate value for the top and bottom diameters for all the samples. Consequently, to account for geometrical fluctuations of the arrays, the top and bottom diameters of the truncated conical-shaped pillars for the S3 array were varied respectively between 300 and 350 nm for the $d_T$ and between 425 and 475 nm for $d_B$. Next, a weighted average determined from a linear-least square fit to normal incidence experimental data (Figure 3c of main text, blue line) was used.

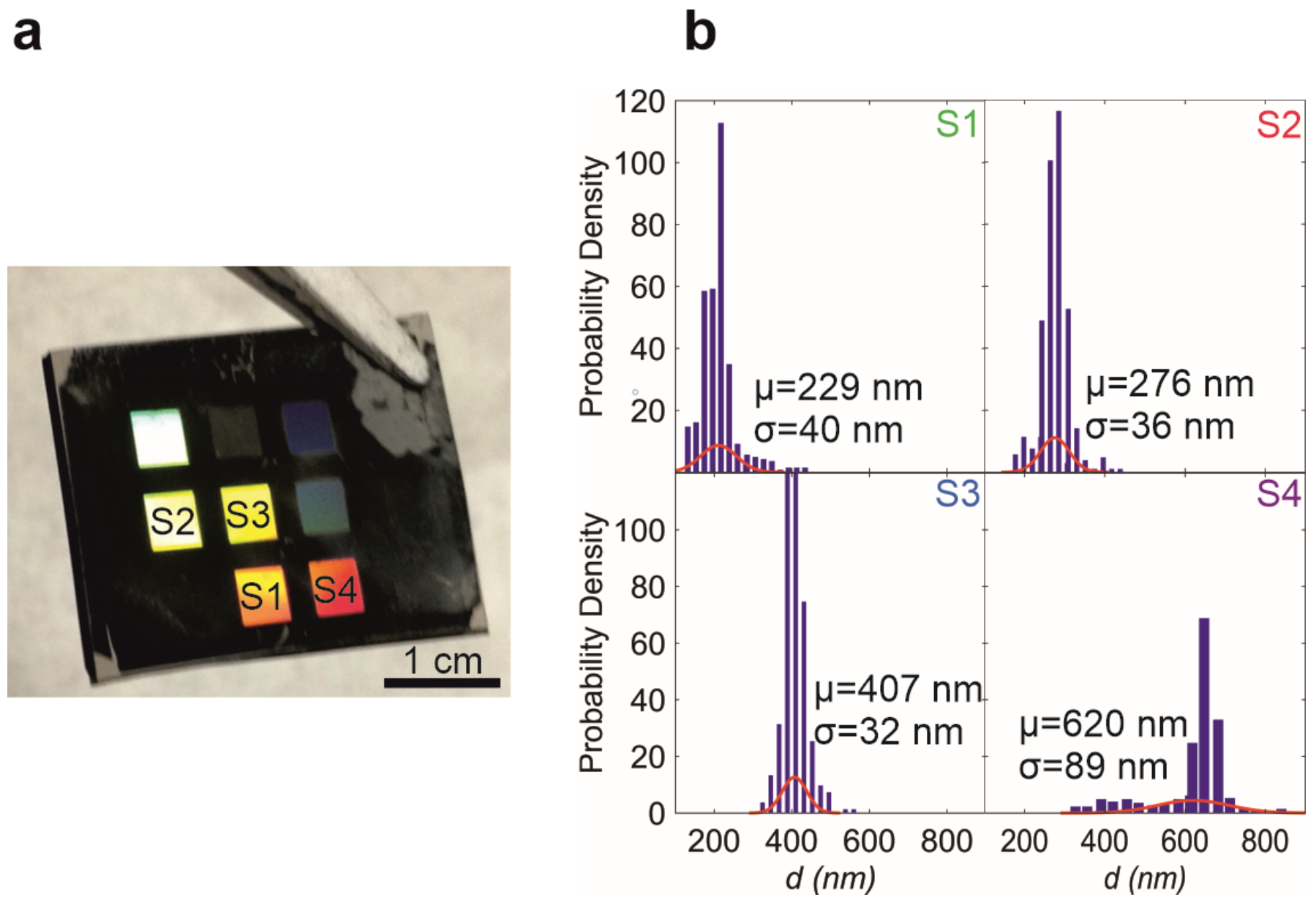

**Figure S5.** (a) Optical micrograph of the four microfabricated NW arrays. Nanostructures Light scattering is clearly visible due to the different diameter of the arrays. (b) Statistical estimation of the diameter distribution of the fabricated NW arrays (S1-4). The diameter is measured from the middle section of the NW. The red curve is a normal distribution fit of the measured diameters, with an $R^2$ higher than 0.98. The diameter was measured from 20x20 μm$^2$ SEM maps and was estimated at the middle of the NW.

## S6. Absorption Measurements.

Optical characterization: the optical band gap of the 1.1 μm-graded GeSn layer was characterized with three independent spectroscopic techniques. To begin with, a combination of VIS spectroscopic ellipsometry and UV-VIS-NIR spectrophotometry techniques were undertaken. SE measured the change of the reflectance for $s$ and $p$ polarization while a direct estimation of the transmission and reflectance spectra were measured for both polarizations. The angle of incidence (AOI) was changed from 60° to 85° with a 5° step so that the spectrophotometry measurement matches those of the SE. Polarization dependant reflectance and transmission were measured with the Varian UMA spectrophotometer. To build the optical model discussed in Section S2, reflectance and transmittance were fitted simultaneously with the ellipsometry parameters $(\Psi, \Delta)$ to extract the dielectric constant. To evaluate the absolute absorption, the Perkin Elmer (PE) scanning UV-VIS-NIR spectrophotometer Lambda 1050 was used. It is a UV/Vis/NIR double beam and double monochromator instrument with two light sources, a deuterium lamp for the UV range and a halogen lamp for the Vis/NIR range. Two lamps are used to achieve the following spectral range - a deuterium lamp produces broadband UV light from 190 to 450nm and a halogen lamp produces broadband light from 450 to 2500 nm. The beam source has a spectral range from 175 to 2500 nm, an incident angle of $\theta_i$=8° and about 17×9 $mm^2$ size. The beam is divided by a beam splitter (BS) into a sample and reference beam (see Fig. 3). The double beam alignment increases the measurement accuracy, allowing to reduce errors due to lamp intensity variations. The detection system is composed of a photomultiplier for the UV/Vis range and a Peltier controlled PbS (Lead (II) Sulfide) detector for the NIR range. However, the model Lambda 1050 incorporates a 125mm integrating sphere (IS) with a Peltier cooled InGaAs (Indium Gallium Arsenide) detector for the NIR range, which provides a better level of sensitivity, resolution and scanning speed. The InGaAs detector was used in this work. The maximum resolution in both models is 0.05 nm in the UV-Vis range and 0.2 nm in the NIR range. The manual calibration is done by measuring first the zero line, $I_0$, which consists of removing any sample from the measuring port of the IS and blocking the incident light to penetrate the IS. The second step consists of putting the calibrated standard on the measuring port, opening the incident light port, and measuring it as the baseline, $I_{100}$. The calibrated standard is a reference mirror of known reflectance, $I_{ref}$, that was calibrated by an external accredited laboratory. Finally, the sample to be measured is situated in the measurement port (inside the sphere) and the sample signal, $I_{sample}^{raw}$, is obtained. The corrected reflectance of the sample, $I_{sample}^{corr}$, is obtained by applying the Eq. (3). During the measurement, the baseline was checked several times during a working day to detect possible drifts.

$$I_{sample}^{T} = I_{ref} \frac{I_{sample}^{raw,T} - I_0}{I_{100} - I_0} \qquad (3)$$

Besides, Diffuse reflectance and transmittance spectra were also recorded in the PE spectrophotometer. When the diffuse reflectance is measured, the specular beam leaves the sphere through an open specular port plug (see Figure S6) The total reflectance is recorded with this port closed. The diffuse transmittance is recorded if the collimated transmitted beam enters a beam dump or a light trap (with 0.01% transmission), and the total transmittance is recorded with the beam dump replaced by the PTFE (polytetrafluoroethylene or Spectralon) reflectance standard. The integrating sphere has been calibrated versus Spectralon diffuse reflectance standards. The reflectance of the NW array was calculated using the equation (4)

$$I_{sample}^{R} = I_{Spectralon}^{0} \frac{I_{sample}^{raw,R} - I_0}{I_{100} - I_0} \qquad (4)$$

where $I_{Spectralon}^{0}$ is the reflectance of the standard (measured by the IS manufacturer), and $I_{sample}^{raw}$, $I_{100}$, and $I_0$ are the reflected signals from the sample, the baseline, and the zero line, respectively. Additionally, spectrally resolved measurements of the total transmittance and reflectance were conducted at room temperature by ultraviolet–visible–near infrared (UV–Vis–NIR) spectroscopy using the IS. In total transmission mode, the NW array was mounted on a well-defined aperture (as shown in Figure S6) and the transmitted light was collected over a large range of angles. Each measurement was calibrated using the Spectralon reflectance standard. For total reflection measurements, the sample was mounted inside the integration sphere at a position opposite of the transmission slit of the aperture and at an 8° incidence angle. For both transmission and reflection measurements, the NW array was facing the incident beam to minimize the influence from the substrate.

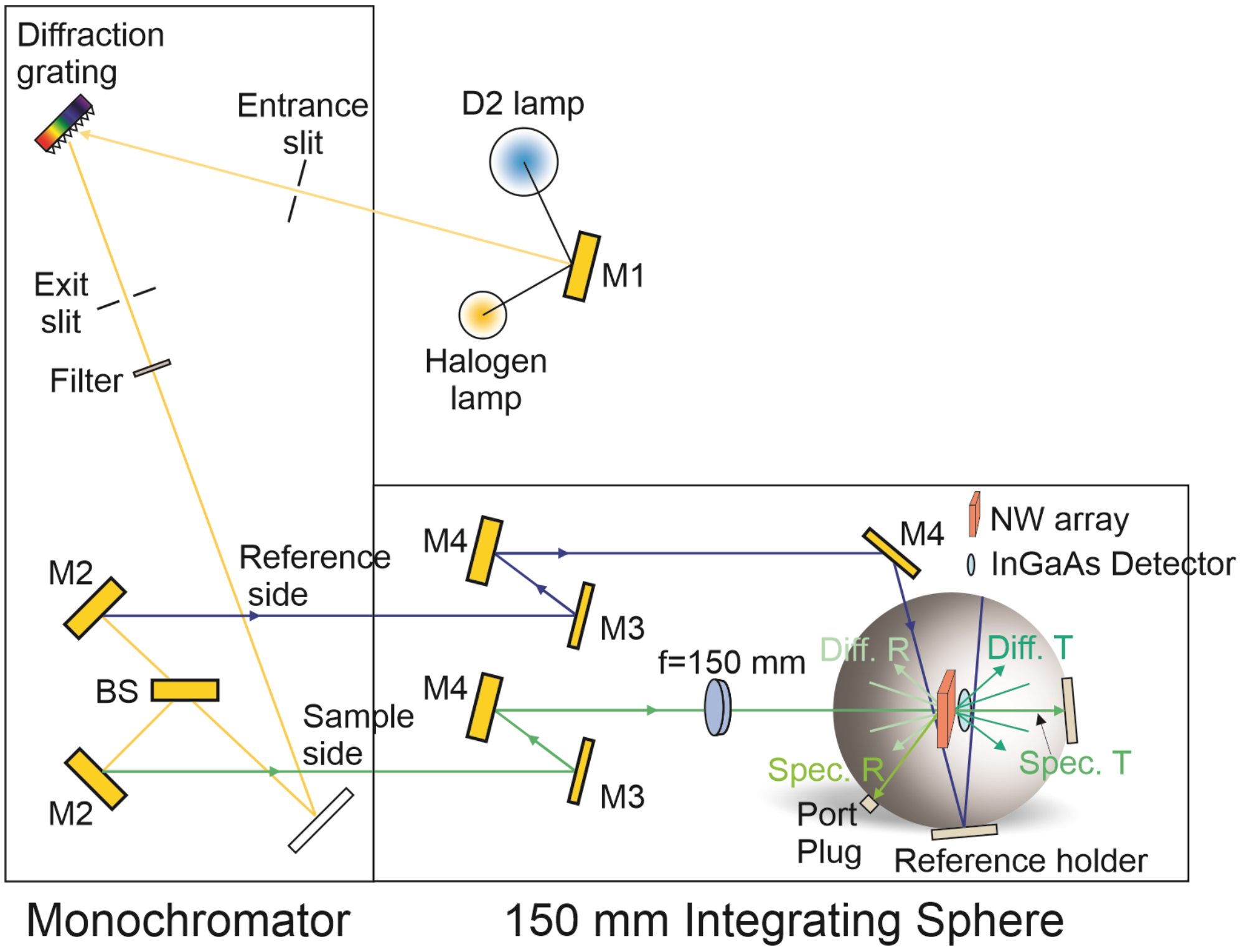

**Figure S6.** Integrating sphere scheme used to measure the absorptance, reflectance and transmittance of the NW arrays. The Halogen lamp source emits between 450 nm and 2500 nm spectral range. Then, the incident light enters a diffraction grating monochromator where each wavelength is sent at a time to the IS. The monochromatic wavelength is then separated by a beam splitter (BS) to a reference and a sample beam, thus the double beam connotation, and sent directly inside the integrating sphere through a series of mirror as shown in the sketch. Finally, the sample beam hits the NW arrays whereas the reference beam hit the PFTE Spectralon and the signal is detected simultaneously with an InGaAs detector. Inserting the sample inside the IS allows for an automatic measurement of the diffuse and specular reflectance and transmittance, which then can be inherently incorporated to quantify the absorptance of the NW arrays.

## S7. Modal Analysis.

Figure S7 shows the effective index for guided modes supported in a cylindrical free-standing GeSn nanowire, as a function of the NW diameter. The mode wavelength in free-space is 2.01 μm, which is very close to the GeSn band-edge emission at room temperature. The effective index for these modes varies between the refractive index of air (1) and the GeSn (4.3489. Near the mode-cut-off, the effective index is close to the air index and the mode is weakly confined within the NW. As the NW diameter increases, the effective index increases, resulting in a better mode confinement.

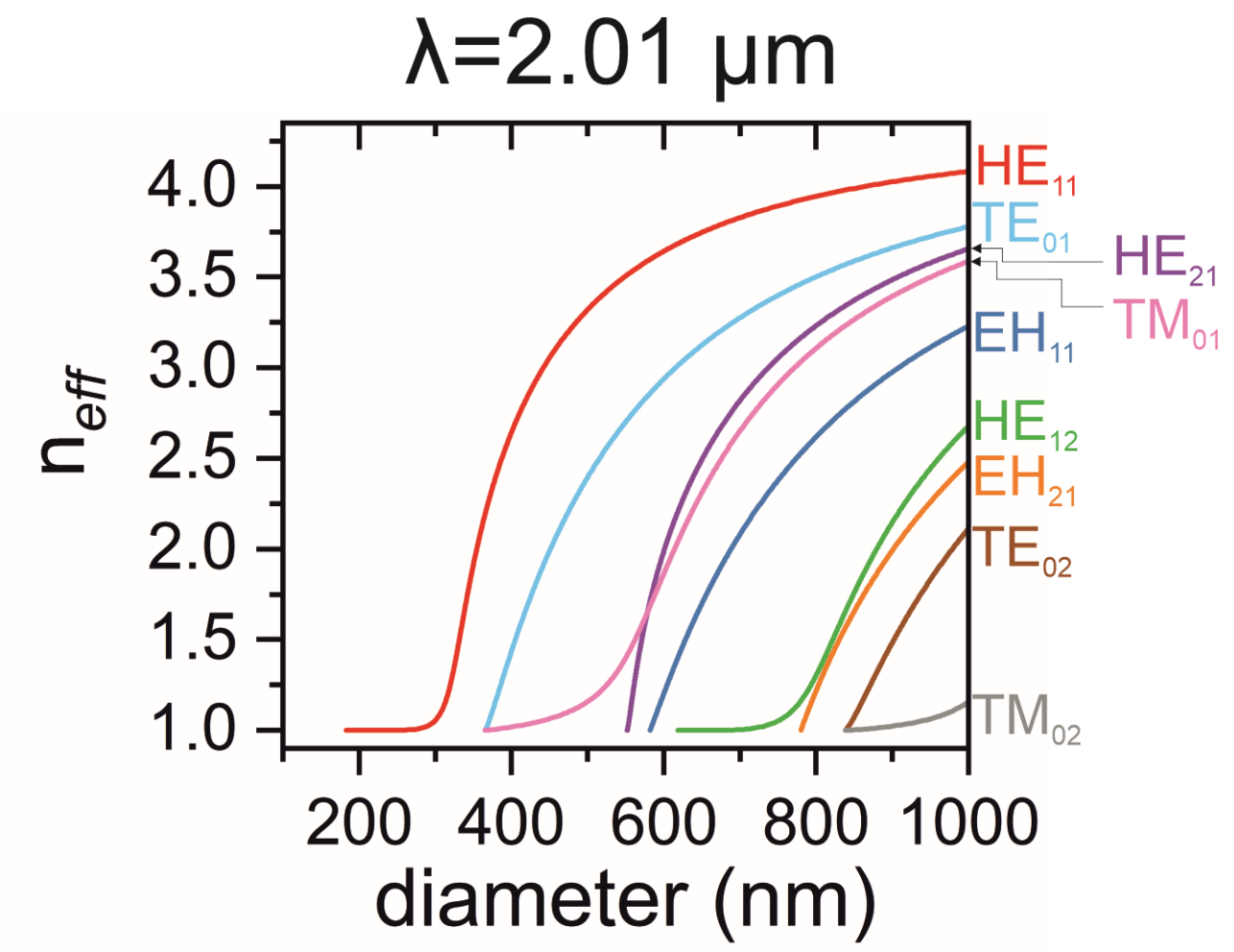

**Figure S7.** Effective index of supported modes versus nanowire diameter for free-standing GeSn NW in air.

The NW array is excited with randomly positioned dipole sources inside the simulation domain. The sources inject a short pulse of radiation to excite a broad range of frequencies. Both TE and TM dipoles sources polarization were excited. Next, the excitation pulse is set between 0.9 and 2.5 μm. Fourier transform analysis of the time signals of the fields reveals the mode frequencies. Figure 3a presents the different modal frequencies obtained for each NW array. The quality of a resonance mode is then defined as $Q = \lambda/\Delta\lambda$, where $\lambda$ and $\Delta\lambda$ are the wavelength and full-width half-maximum (FWHM), respectively, of the resonance band extracted from a Lorentzian fit to the observed peaks, indicated by arrows in Figure 3a.

Incident light couples to $HE_{1n}$ modes more favourably compared to other NW modes, namely $TE_{0n}$, $TM_{0n}$, $HE_{mn}$ ($m > 1$), $EH_{mn}$ ($m > 1$), which, barring specific cut-off conditions for modes, all exist within the NW waveguide. This can be understood by analysing the symmetry properties of these modes and how closely their E-field pattern resembles that of the incident plane wave [6,7] The closer the resemblance, the more coupling overlap the mode will have with the incident light. A plane wave, with E-field polarized in the *x* or *y* direction, is anti-symmetric under reflection along *y-z* or *x-z* plane that axially bisects the NW (cylindrical waveguide). As a result, the plane wave is circularly asymmetric and has angular mode index $m = 1$ (π rotation returns to the same value). This rule out excitation of all modes except for $m = 1$ group of modes, namely: $HE_{1n}$ and $EH_{1n}$. Between these two groups, the transverse E-field pattern of the $HE_{1n}$ modes closely resembles that of the plane wave, inside and outside the NW core, with the E-field pattern of the $HE_{11}$ mode resembling it the most. Therefore, the $HE_{1n}$ family of modes are the dominant excited modes in vertical NWs. [8] A single, isolated nanowire can show guided $HE_{mn}$, $EH_{mn}$, $TE_{0n}$, and $TM_{0n}$ modes[9]. Coupling between two modes is allowed only if their difference in *m* value is even. Hence, the incident light can excite only $HE_{mn}$ and $EH_{mn}$ modes with odd *m*, and excitation of $TE_{0n}$ and $TM_{0n}$ modes is not possible. We solve for 200 allowed eigenmodes of the NW array (the incidence angle, the incident polarization, and geometrical symmetry of the array determine which modes can be excited with the incident light), we find that with decreasing wavelength, each array mode converges to one specific fundamental mode $HE_{11}$(Figure S8b). For all modes in the NW array, except for the fundamental $HE_{11}$, $n_{\alpha}$ goes toward the medium effective index of 1 with increasing wavelength, which corresponds to weakly confined modes.

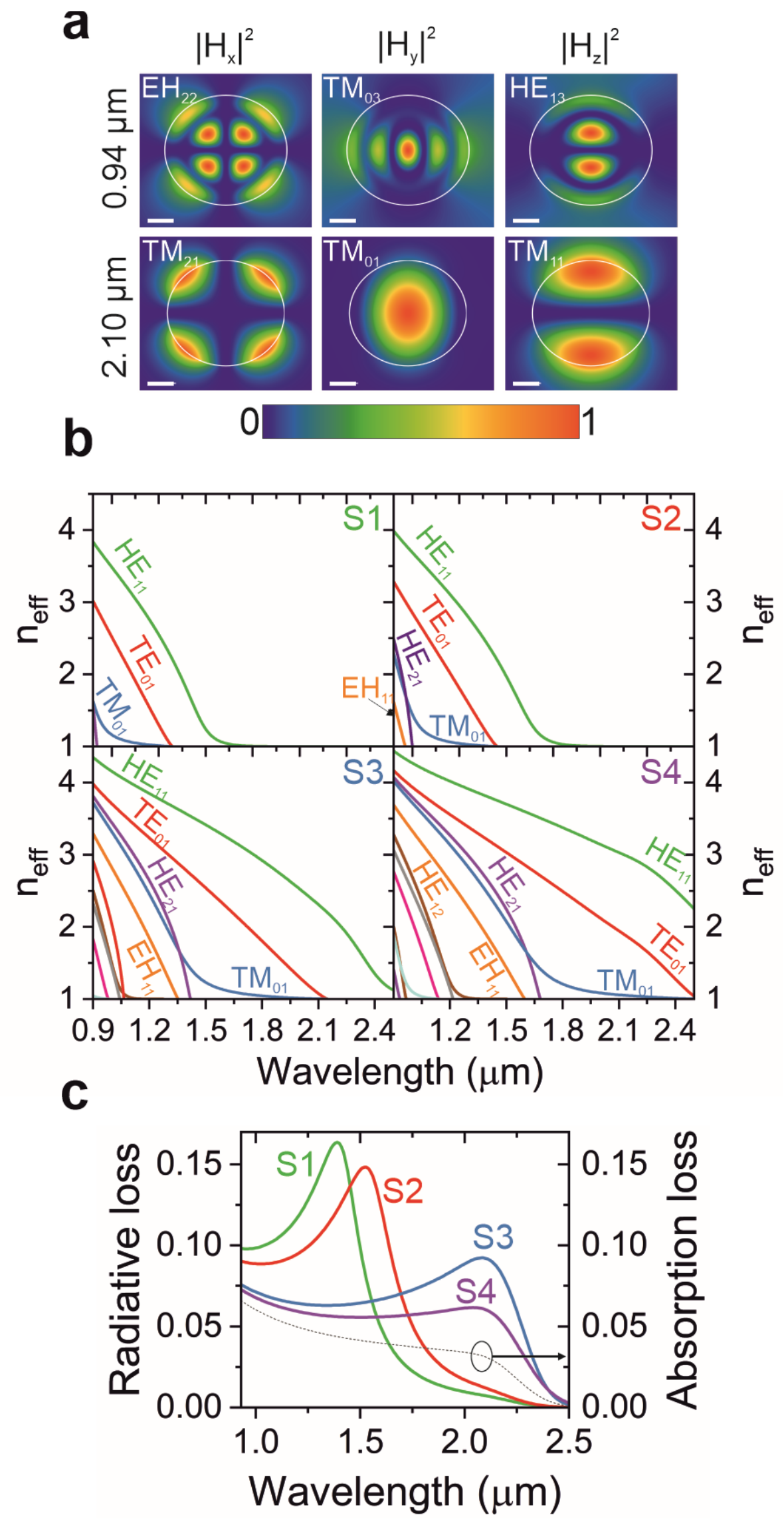

**Figure S8. (a)** Magnetic field intensity profiles for the S4 NW array with a diameter estimated at a height of 0.45 µm at 0.94 µm ($EH_{22}$ mode for x-component; $TM_{03}$ mode for the y-component and $HE_{13}$ for the z-component) and at 2.10 µm ($TM_{21}$ mode for x-component; $TM_{01}$ mode for the y-component and $TM_{11}$ for the z-component). **(b)** Dispersion of the optical modes in the S1-4 NW arrays. The diameters are evaluated at mid-height of the NWs for all the arrays. Only propagating modes of the NW arrays are shown. Overlaid are the 5 first modes for each structure. **(c)** The radiative loss as well as the absorption loss for each NW array is calculated.